\documentclass[pre, reprint, superscriptaddress, amsmath, amssymb, aps]{revtex4-2}

\usepackage{graphicx}
\usepackage{dcolumn}
\usepackage{bm}
\usepackage[protrusion=true,expansion=true]{microtype}
\usepackage[normalem]{ulem}
\usepackage{soul}
\usepackage{xcolor}
\usepackage{kotex}
\usepackage{hyperref}
\usepackage{cleveref}
\usepackage{comment}

\def\min{\textrm{min}}

\begin{document}

\preprint{APS/123-QED}

\title{Heterogeneous Network Topology Induces the Widom Line}

\author{Cook Hyun Kim}
\affiliation{CCSS, KI for Grid Modernization,
Korea Institute of Energy Technology, Naju, Jeonnam 58330, Korea}

\author{B. Kahng}
\email{bkahng@kentech.ac.kr}
\affiliation{CCSS, KI for Grid Modernization,
Korea Institute of Energy Technology, Naju, Jeonnam 58330, Korea}

\date{\today}

\begin{abstract}
The Widom line, initially identified as a crossover line between liquid-like and gas-like behavior in water and supercritical fluids, separates these two types of behavior. Here, we show that an analogous line arises in spin models on scale-free networks as a consequence of degree heterogeneity, which we analyze using the annealed network approximation. For the Ashkin--Teller and Invisible Potts models, the Widom line exists within a finite range of the degree exponent. It separates two distinct ordered regimes$-$distributed spin alignment and hub-dominant alignment$-$while also giving rise to a supercritical-like state where the two alignments become indistinguishable. These results demonstrate that degree heterogeneity alone can generate mesoscopic crossovers beyond conventional phase-transition theory, opening new directions for understanding and controlling collective dynamics in complex networks. 
\end{abstract}

\maketitle

\section{Introduction}
Phase transitions and critical phenomena are central topics in statistical physics~\cite{fisher1967theory,stanley1971phase,wilson1975renormalization, cardy1996scaling}, arising from macroscopic singularities generated by microscopic correlations. Conventional transitions manifest themselves as either divergences in response functions or discontinuous jumps in the order parameter. However, beyond these phase boundaries, additional crossover phenomena can occur in the form of the Widom line~\cite{widom1965surface}, along which thermodynamic response functions such as susceptibility exhibit maxima without true singularities~\cite{cockrell2021transition,li2024thermodynamic}.

The Widom line was first identified in supercritical fluids~\cite{simeoni2010widom, maxim2019visualization,abascal2010widom,gallo2014widom}, where it marks a crossover from liquid-like to gas-like behavior in the absence of an actual phase boundary. Similar crossovers appear in spin-glass models~\cite{binder1986spin,mezard1987spin} with competing interactions, associated with mesoscopic reorganization and amplified fluctuations, and in neural systems~\cite{destexhe2021there,tian2022theoretical,kinouchi2020mechanisms,fosque2021evidence,fosque2022quasicriticality}, where critical-like crossovers emerge without genuine phase transitions. In all these cases, Widom lines originate from complex microscopic interactions, whether they are competing couplings, frustration, or many-body effects.

Here, we show that Widom lines can arise solely from network topology, without invoking complex microscopic interactions. We consider scale-free (SF) networks~\cite{barabasi1999emergence,albert2002statistical,newman2003structure} with degree distribution $P_d(k)\sim k^{-\lambda}$, where $k$ is the degree and $\lambda$ the degree exponent. Such networks, ubiquitous in complex systems, contain a few highly connected hubs. Although hub effects are known to shape various critical behaviors~\cite{leone2002ferromagnetic,bianconi2002mean,herrero2004ising,lee2009critical,dorogovtsev2002ising,dorogovtsev2008critical}, their role in generating mesoscopic crossover phenomena has not been explored. As a first step, we focus on degree heterogeneity by adopting the annealed network approximation, which isolates its effect by removing degree--degree correlations, clustering, and higher-order motifs. The key ingredient is not the spin dynamics, but the degree heterogeneity itself: hubs impose local order while peripheral nodes remain disordered, generating two competing ordered states. Analyzing the Ashkin--Teller (AT)~\cite{ashkin1943statistics, kadanoff1971some, fan1972symmetry, kohmoto1981hamiltonian} and Invisible Potts (IP) models~\cite{tamura2008first,tamura2010phase,krasnytska2023potts,kim2024entropy}, we find that these two states---globally distributed and hub-dominant alignment---are separated by a Widom line controlled entirely by $\lambda$.

\section{Model and Formalism}
To establish the universality of degree-heterogeneity-induced Widom lines, we analyze two representative spin models: the AT~\cite{jang2015ashkin,kim2021link,kim2025ashkin} and the IP~\cite{sarkanych2022potts} model on the SF network.

As a first step toward understanding how network topology generates Widom lines, we adopt the annealed network approximation~\cite{bianconi2002mean,jang2015ashkin,kim2021link}, replacing the adjacency matrix with its degree-sequence average,
\begin{equation}
    \mathcal{A}_{ij} \;\longrightarrow\; \frac{k_i k_j}{N\langle k \rangle}.
\end{equation}
This removes degree correlations, clustering, and higher-order motifs while preserving the heterogeneous degree distribution, allowing us to attribute any emergent phenomenon solely to degree heterogeneity.

The AT model involves coupled multi-spin interactions, with each node hosting two Ising spins $(s_i,\sigma_i=\pm1)$, and the Hamiltonian is expressed as
\begin{equation}
-\beta\mathcal{H} = K_2 \sum_{\langle i,j\rangle} (s_is_j + \sigma_i\sigma_j) + K_4 \sum_{\langle i,j\rangle} s_i\sigma_i s_j\sigma_j ,
\end{equation}
where $\beta=1/T$, $K_2=J_2/T$, and $K_4=J_4/T$, with $J_2=1$ and $x\equiv J_4/J_2$ controlling the strength of the four-spin coupling. Under the annealed approximation, the Hamiltonian becomes
\begin{align}
    -\beta\mathcal{H}_{\rm ann} 
    & = \frac{K_2}{2 N\langle k\rangle}\sum_{i,j} k_ik_j \left(s_is_j + \sigma_i\sigma_j\right) \nonumber \\
    & + \frac{K_4}{2 N\langle k\rangle} \sum_{i,j} k_ik_j s_i\sigma_i s_j\sigma_j.
\end{align}
The system exhibits three phases: the Baxter phase ($m>0,\,M>0$), the $\langle\sigma s\rangle$ phase ($m=0,\,M>0$), and the paramagnetic phase ($m=M=0$), where the order parameters are defined as
\begin{align}
m &\equiv \frac{1}{N\langle k \rangle}\sum_i m_i k_i, & m_i &= \langle s_i \rangle = \langle \sigma_i \rangle, \\
M &\equiv \frac{1}{N\langle k \rangle}\sum_i M_i k_i, & M_i &= \langle \sigma_i s_i \rangle,
\end{align}
with $k_i$ denoting the degree of node $i$. Throughout this work, $\langle \cdot \rangle$ denotes an ensemble average, while $\langle k \rangle$ represents the mean degree of the network. The nature of the phase transition depends sensitively on $x$: it is continuous for $x\ll1$, discontinuous near $x\simeq1$, and successive continuous transitions for $x\gg1$.

The IP model captures the competition between interaction and entropy. Each spin has $q$ visible states and $r$ invisible states, with the Hamiltonian,
\begin{equation}
-\beta\mathcal{H} = K \sum_{\langle i,j\rangle} \sum_{\alpha=1}^{q} \delta_{s_i,\alpha}\delta_{s_j,\alpha},
\end{equation}
where $K=J/T$. Ferromagnetic interactions act only among visible states, whereas invisible states contribute purely entropic weight. Under the annealed approximation, the Hamiltonian becomes
\begin{equation}
-\beta\mathcal{H}_{\rm ann} = \frac{K}{2 N\langle k\rangle} \sum_{i,j} \sum_{\alpha=1}^{q} k_i k_j \delta_{s_i,\alpha} \delta_{s_j,\alpha}.
\end{equation}
The order parameters are defined as
\begin{align}
    m &\equiv \frac{1}{N\langle k \rangle}\sum_i m_i k_i, & m_i &= \langle \delta_{s_{i}, \alpha=1}-\delta_{s_{i}, \alpha=2} \rangle \\
    m_r &\equiv \frac{1}{N\langle k \rangle}\sum_i m_{r,i} k_i, & m_{r,i} &= \sum_{\alpha \ne [1,2]}\langle \delta_{\alpha,i}\rangle
\end{align}
which characterize visible alignment and invisible-state occupation, respectively. The system supports Ferro ($m>0,\,m_r>0$) and Para ($m=0,\,m_r>0$) phases, with the transition type controlled by the parameter $x=r/q$.

Under this approximation, the scale-free topology introduces a fundamental asymmetry: hubs, due to their high connectivity, impose local spin order even as the periphery remains disordered. This mechanism produces two competing ordered states: distributed ordering and hub-dominant ordering, whose competition is governed purely by $\lambda$.

\begin{figure}[!t]
\centering
\includegraphics[width=1.0\linewidth]{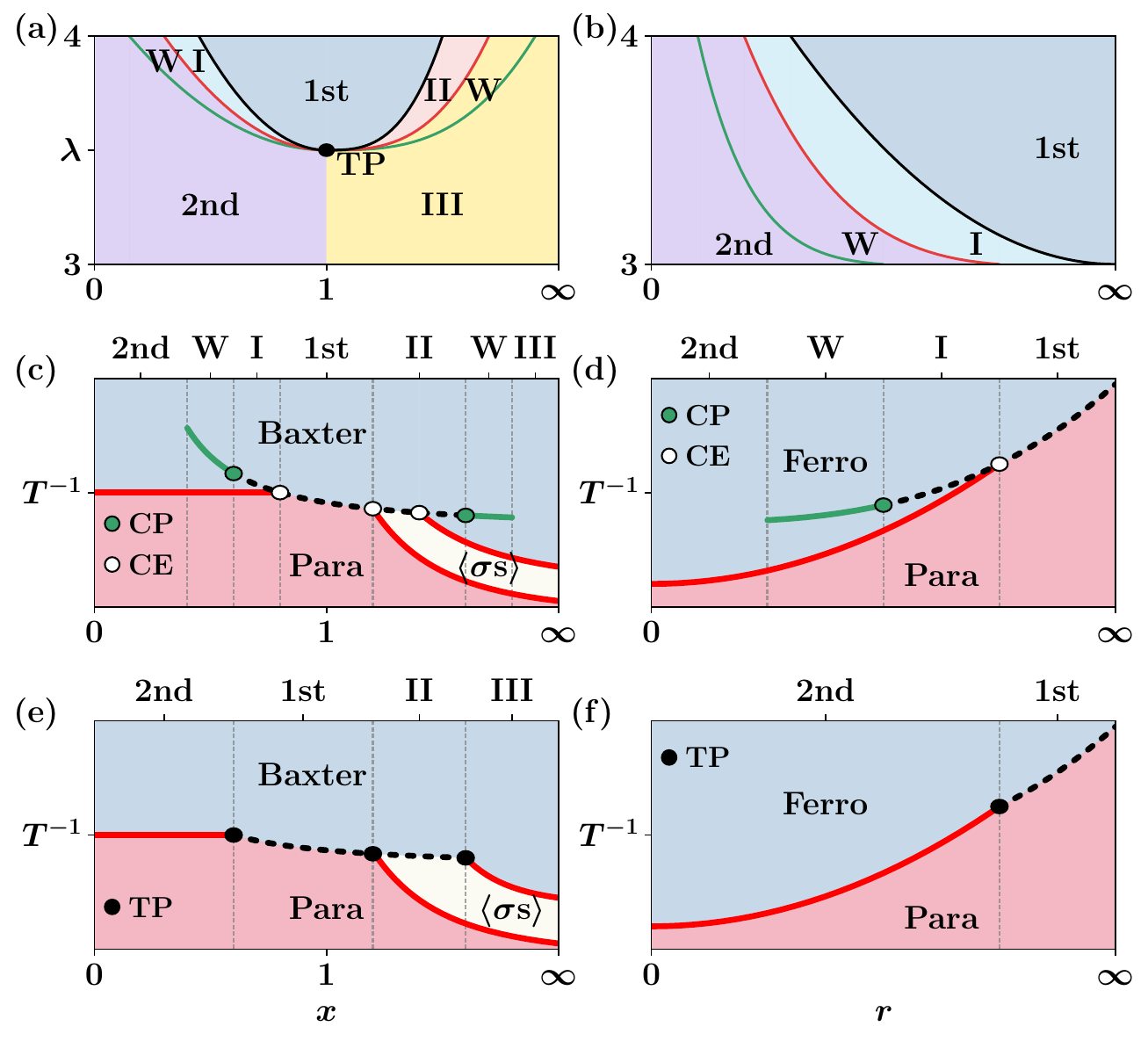}
\caption{Schematic phase diagrams based on mean-field analysis for the AT model (a,c,e) and the IP model (b,d,f). 
(a) $(x,\lambda)$ and (b) $(r,\lambda)$ planes on SF networks; 
(c) $(x,T^{-1})$ and (d) $(r,T^{-1})$ planes on SF networks; 
(e) $(x,T^{-1})$ plane on homogeneous networks (regular lattices or Erdős--Rényi graphs), 
where CP, the Widom line, and hub-dominant order are absent. 
Acronyms: CP (critical point), CE (critical endpoint), TP (tricritical point). 
Black dashed, red solid, and green solid lines denote discontinuous, continuous, and Widom lines, respectively. 
Widom lines (green) emerge only in heterogeneous networks.}
\label{fig:fig1}
\end{figure}

\subsection{Numerical Methods}

To verify that degree heterogeneity alone drives the observed crossover phenomena, Monte Carlo simulations were performed on the annealed Hamiltonian $\mathcal{H}_{\rm ann}$ using the Metropolis algorithm. The parameters used in Fig.~\ref{fig:fig1}, Figs.~\ref{fig:figS5} and~\ref{fig:figS6} are $N = 10^6$, $k_{\min} = 1$, and $\lambda=3.9$ ($4.8$) for the AT (IP) model, both in the intermediate regime. Starting from a fully aligned initial condition, all spins are updated $2.25 \times 10^4$ times to reach equilibrium, and this process is repeated $1.225 \times 10^3$ times to obtain the ensemble. Further details are provided in the Supplemental Material~\cite{SuppMat}, which includes Refs.~\cite{bianconi2002mean, jang2015ashkin, kim2021link}.

\section{Main Result}

The degree exponent $\lambda$ serves as the principal control parameter of degree heterogeneity, dictating which ordered states emerge and how they compete. A small $\lambda$ produces highly skewed degree distributions dominated by hubs, whereas a large $\lambda$ yields nearly distributed connectivity approaching a regular lattice.

These structural differences define four regimes of phase behavior. For $2<\lambda<3$, extreme heterogeneity produces dominant hubs that suppress thermal fluctuations and enforce global spin alignment even at high temperatures. In this regime, strong hub influence renders the distributed and hub-dominant orderings indistinguishable, resulting in a supercritical-like state where the two mechanisms merge. Beyond $\lambda=3$, there exist two characteristic values of $\lambda$: $\lambda_d$ and $\lambda_u$, whose derivations are presented in the Supplemental Material (SM)~\cite{SuppMat}. For $3<\lambda<\lambda_{d}$, the hub influence is relatively weak, so that a phase transition occurs at a finite temperature; the supercritical-like state is nevertheless preserved. In the intermediate range $\lambda_{d}<\lambda<\lambda_{u}$, hubs no longer impose global alignment but still generate local order in their neighborhoods, leading to competition between hub-dominant and distributed ordering. This coexistence produces phase diagrams featuring both discontinuous and continuous transitions, giving rise to Widom lines that mark smooth crossovers between ordering modes. For $\lambda>\lambda_{u}$, the SF network becomes effectively homogeneous: hub asymmetries vanish, hub-dominant ordering disappears, and intraphase transitions terminate.

The phase diagrams of the AT and IP models corroborate this classification, showing that the Widom line (green curves in Fig.~\ref{fig:fig1}) arises exclusively in the intermediate regime $\lambda_d < \lambda < \lambda_u$. This result indicates that degree heterogeneity acts as a key governing factor for crossover behavior in complex systems.

\begin{figure}[!t]
\centering
\includegraphics[width=1.0\linewidth]{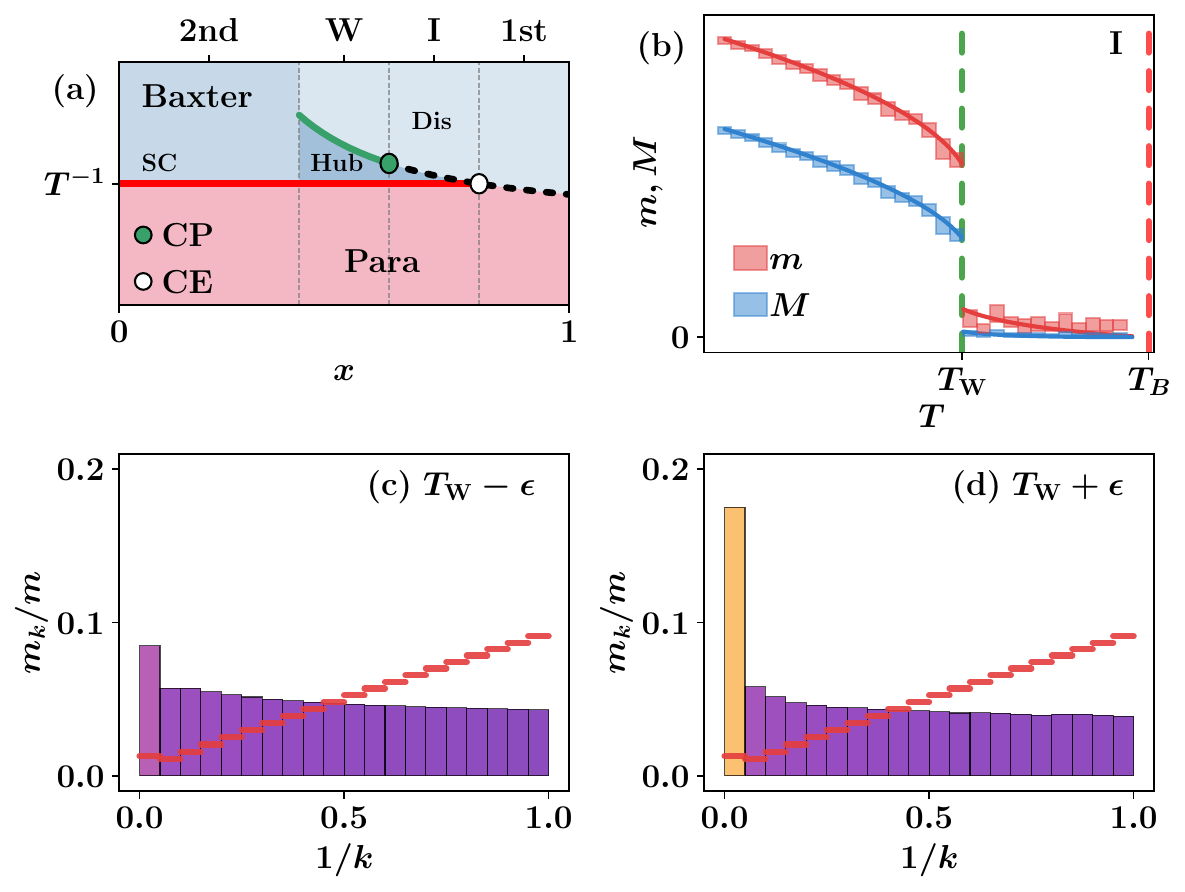}
\caption{(a) Schematic phase diagram based on mean-field analysis, showing the subdivision of the Baxter phase for the $x<1$ regime of the AT model. \textbf{Dis}, \textbf{Hub}, and \textbf{SC} denote distributed, hub-dominant, and supercritical orderings, respectively. (b) Monte Carlo results for the order parameters $m$ and $M$ versus temperature in the I regime of (a) with $x=0.72$ and $\lambda=3.9$; colored boxes indicate the mean $\pm$ one standard deviation, and solid lines indicate mean-field predictions. Monte Carlo simulations are performed on the annealed network Hamiltonian $\mathcal{H}_{\rm ann}$, so that fluctuation effects are fully captured. (c,d) Degree-resolved magnetization $m_k$ versus $1/k$ for (c) distributed and (d) hub-dominant states. The interval $1/k \in [0,1]$ is divided into 20 equal bins; for each bin we compute $m_k = \sum_{i \in k} m_i k_i / (N \langle k \rangle)$, with $\sum_k m_k = m$. Each bar shows the contribution of one bin to the total magnetization, while red staircases indicate the degree's contribution to gauge the expected contributions. $T_{\rm w} \approx 2.0974$ denotes the crossover temperature and $\epsilon$ a small parameter. Results for the $x>1$ regime of the AT model and for the IP model are provided in the Supplemental Material (Figs.~\ref{fig:figS5} and \ref{fig:figS6}).}
\label{fig:fig4}
\end{figure}

\begin{figure}[!t]
\centering
\includegraphics[width=1.0\linewidth]{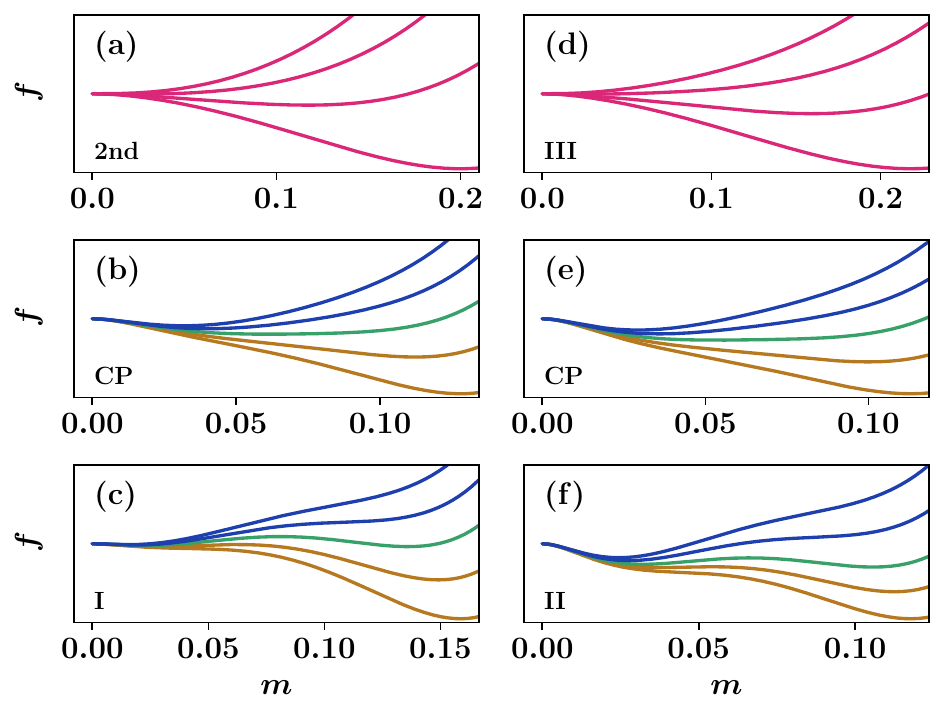}
\caption{Landscapes of the Ginzburg--Landau free energy for different $x$ regimes in Fig.~\ref{fig:fig1}(c). Amber and navy curves represent distributed and hub-dominant ordered states, respectively. Green curves indicate continuous or crossover transitions between them. Pink curves denote a supercritical-like state where the two orderings become indistinguishable. Results for the IP model are presented in the Supplemental Material (Figs.~\ref{fig:figS7}).}
\label{fig:fig3}
\end{figure}

\subsection{Microscopic Spin Configurations}

The degree-resolved magnetization patterns in Fig.~\ref{fig:fig4} highlight the contrast between the two ordering modes: panel (c) shows distributed ordering, whereas panel (d) displays hub-dominant ordering.

Heuristic analysis also supports this distinction. Near the critical temperature, where the order parameter is small ($m \ll 1$), the local magnetization of a node with degree $k$ obeys
\begin{align}
m_i \sim \tanh\Big(\dfrac{m k_{i}}{T}\Big) \sim
\begin{cases}
\dfrac{m k_{i}}{T} \ll 1, & k_{i} \ll \dfrac{T}{m}, \\[12pt]
1, & k_{i} \gg \dfrac{T}{m},
\end{cases}
\end{align}
indicating that low-degree nodes (peripherals) remain largely unaligned, whereas high-degree nodes (hubs) maintain alignment, producing hub-dominant ordering.

At low temperatures, where the order parameter approaches unity ($m \sim 1$), the local magnetization satisfies
\begin{align}
m_i \sim \tanh\Big(\dfrac{m k_{i}}{T}\Big) \sim 1 \quad \text{for all } k_{i},
\end{align}
indicating uniform alignment across all nodes. In this regime, the network structure does not play a role, and the system exhibits distributed ordering.

\begin{figure}[!t]
\centering
\includegraphics[width=1.0\linewidth]{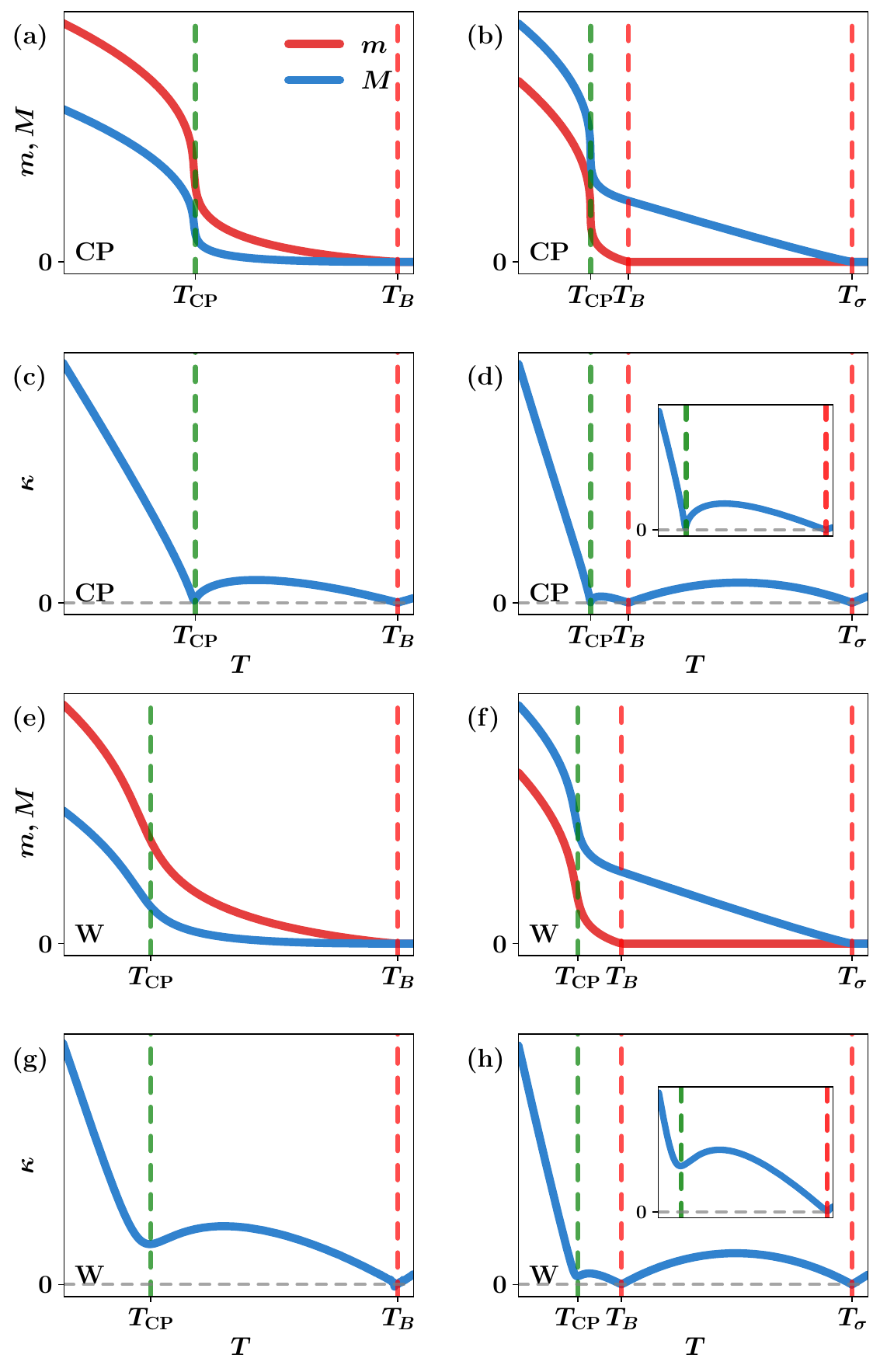}
\caption{Order parameters $m$ and $M$ and free-energy curvature $\kappa$ as functions of temperature for $x<1$ (a,c,e,g) and $x>1$ (b,d,f,h) in the AT model. Panels (a$-$d) correspond to $x_{\rm CP}$ in Fig.~\ref{fig:fig1}(c), while (e$-$h) are taken from region W in Fig.~\ref{fig:fig1}. $T_{\rm CP}$ and $T_{\rm B}$ denote the critical-point temperature and the transition temperature between the paramagnetic and Baxter phases, respectively. Other phase transitions and the IP model are presented in the Supplemental Material (Figs.~\ref{fig:figS1}--\ref{fig:figS4}).}
\label{fig:fig2}
\end{figure}

\noindent{\it Widom Line---}
Whether these two ordered states constitute thermodynamically distinct phases or merely quantitative variations depends on the presence of transitions between them. Depending on control parameters, degree heterogeneity, temperature, and interlayer coupling in the AT model, or the number of invisible states in the IP model, the system may exhibit discontinuous, continuous, or crossover transitions, or none at all.

The nature of these transitions is encoded in the Ginzburg--Landau (GL) free energy $f$ and its curvature $\kappa \equiv f_{mm}$, the second derivative with respect to $m$. This curvature provides the central diagnostic: discontinuous transitions correspond to abrupt jumps in $\kappa$, continuous transitions to singular behavior with $\kappa=0$, crossover behavior to finite $\kappa$ with sharp local minima, and absence of transitions to smooth monotonic variations of $\kappa$ without pronounced features.

To demonstrate this diagnostic, we begin with $x$ close to 1, where the crossover effects are strongest. Here, the free energy develops two distinct minima associated with hub-dominant and distributed orderings, producing discontinuous transitions within the ordered regime [Fig.~\ref{fig:fig3}(c,f)]. As $x$ deviates from unity, these minima flatten and eventually merge at a critical point (CP) [Fig.~\ref{fig:fig3}(b,e)], where
\begin{equation}
\kappa = \left. \dfrac{\partial^2 f}{\partial m^2} \right|_{m=m^{*}, M=M^{*}} = 0,
\end{equation}
signaling divergent susceptibility ($\chi \sim \kappa^{-1}$) and the onset of a second-order transition [Fig.~\ref{fig:fig2}(c,d)].

Moving further from $x=1$, the order parameters $m$ and $M$ vary smoothly with temperature, without singularities or discontinuities [Fig.~\ref{fig:fig2}(e--h)]. In this regime, $\kappa$ remains finite, but develops a pronounced local minimum that approaches, yet never reaches, zero [Fig.~\ref{fig:fig2}(g,h)]. This minimum marks the point of the maximal thermodynamic response without singularity, defining the signature of a crossover. Consequently, at this temperature, susceptibility and correlation length peak, while $m$ and $M$ undergo rapid changes as the system reorganizes between hub-dominant and distributed orderings. This behavior is directly analogous to the Widom line in fluids, which identifies the locus of maximal response in the absence of an actual phase transition.

For larger deviations from $x=1$, $\kappa$ varies monotonically and no longer distinguishes between the two states. The two phases merge into one, entering a supercritical-like regime in which the GL free-energy landscape exhibits only a single local minimum, confirming the absence of competing ordered states [Fig.~\ref{fig:fig3}(a,d)].

\subsection{Mean-Field Analysis}
The complete theoretical picture emerges from the explicit GL expansion, which provides general explanatory power for competing ordering modes and their crossover behavior. Here we focus on the regime $3<\lambda<4$, where Widom lines are most prominently observed (see the SM~\cite{SuppMat} for detailed derivations).
For $x<1$,
\begin{align}
    f_{x<1}
    & \sim \left(\dfrac{\langle k \rangle}{T} - \dfrac{\langle k^{2} \rangle}{T^{2}}\right)m^2
    - 2C(\lambda)\left(\dfrac{m}{T}\right)^{\lambda-1} \cr
    & \quad
    -\dfrac{x[D_{<}(\lambda)]^2}
    {2T\left(\langle k \rangle - x\langle k^2 \rangle/T\right)} \left(\dfrac{m}{T}\right)^{2(\lambda-2)},
    \end{align}
while for $x>1$,
\begin{align}
    f_{x>1}
    & \sim \left(\dfrac{\langle k \rangle}{T} - \dfrac{\langle k^{2} \rangle}{T^{2}}
    + C_{>}(\lambda)\right)m^2 - 2C(\lambda)\left(\dfrac{m}{T}\right)^{\lambda-1} \cr
    & \quad
    -\dfrac{x[D_{>}(\lambda)]^2}
    {2T\left(\langle k \rangle - x\langle k^2 \rangle/T\right)} \left(\dfrac{xM}{T}\right)^{2(\lambda-4)} \left(\dfrac{m}{T}\right)^4.
\end{align}
The quadratic term, together with the $C(\lambda)$ term, generates a small local minimum, producing hub-dominant ordering. In contrast, the $D(\lambda)$ term and higher-order terms (quartic and beyond) generate a large local minimum, producing distributed ordering and thereby establishing competition between the two mechanisms. Near $x=1$, $D(\lambda)$ becomes sufficiently strong for the distributed ordering it induces to compete with hub-dominant ordering, allowing multiple local minima and giving rise to the Widom line. Far from unity, $D(\lambda)$ weakens relative to $C(\lambda)$, restoring a single minimum landscape and leading to a supercritical-like regime in which hub-dominant and distributed orderings become indistinguishable.

Having established the theoretical framework, we now specify the parameter regimes in which competing states and the Widom lines arise (see the Supplemental Material for details). For the AT model, the crossover occurs in $\lambda_d < \lambda < \lambda_{u}$. The lower bound, $\lambda_d \approx 3.50$, marks the threshold below which hub dominance overwhelms the system, rendering the two states indistinguishable. The upper bounds, $\lambda_{u} \approx 7.16$ for $x<1$ and $\lambda_{u} \approx 7.21$ for $x>1$, signal the disappearance of hub dominance, leaving distributed ordering as the only stable state. The IP model shows analogous behavior: distributed ordering competes with hub-dominant ordering, and Widom lines emerge for $3 < \lambda < 8.6$. Thus, neither multispin couplings (AT) nor entropy--interaction competition (IP) alter the degree-heterogeneity-driven mechanism. Both models confirm that Widom-line formation requires intermediate degree heterogeneity: hubs must be strong enough to sustain a hub-dominant state, yet not so strong as to merge it with a distributed state, nor so weak as to eliminate it. This balance defines the regime where degree heterogeneity and interactions conspire to generate crossover phenomena, establishing Widom lines as intrinsic to SF networks rather than as artifacts of specific spin dynamics.

Although macroscopic trends are universal, microscopic details differ. The AT model distinguishes between the $x<1$ and $x>1$ regimes with slightly different $\lambda_u$ values, while the IP model operates over a broader range. The degree-resolved spin patterns also diverge significantly between models (Figs.~\ref{fig:fig2} and~\ref{fig:figS5}--\ref{fig:figS6}). This duality, in which the same degree-heterogeneity-driven mechanism manifests itself through distinct microscopic arrangements, suggests that degree heterogeneity can generate a broad spectrum of crossover behaviors. Conversely, it implies that diverse crossovers across systems may ultimately share a common origin: the interplay of degree heterogeneity and competing ordering tendencies.

\section{Conclusion}

This work establishes a framework linking degree heterogeneity to the emergence of Widom lines and the diversity of ordering transitions. By tuning the degree exponent $\lambda$, we demonstrate that degree heterogeneity gives rise to two distinct ordering regimes: distributed and hub-dominant, as well as a supercritical-like mixed state that spans the full spectrum of phase transitions. This finding identifies degree heterogeneity as a fundamental control parameter for crossover phenomena, complementing previous studies that have examined the effects of network connectivity on equilibrium and dynamical properties.

Although our analysis focused on equilibrium spin systems, the degree-heterogeneity-induced crossover mechanism likely extends to nonequilibrium dynamics as well. For example, random walks on SF networks exhibit similar crossover behaviors~\cite{hwang2012effective,hwang2012first}. The return probability to the starting node shows a crossover from slow to fast decay with time, where the crossover time grows with the degree of the starting node. This trapping effect arises from multiple pathways around hubs, which induce back-and-forth motion near them. More broadly, identifying Widom lines in complex networks may provide a unifying framework for understanding crossover phenomena in various nonequilibrium systems, including social~\cite{galam2008sociophysics,castellano2009statistical} and biological~\cite{mora2011biological,Bialek2014} networks.

Our analysis adopts the annealed network approximation, which isolates degree heterogeneity---the most fundamental structural feature of scale-free networks---by removing degree--degree correlations, clustering, and higher-order motifs, thereby establishing degree heterogeneity as a sufficient condition for the Widom-line formation. However, real networks possess richer internal structure, and how these additional features modify the crossover behavior remains an important open question. Whether assortative mixing or clustering shifts the Goldilocks zone, or whether hypergraphs~\cite{bretto2013hypergraph,battiston2020networks} and multiplex networks~\cite{boccaletti2014structure,kivela2014multilayer,bianconi2018multilayer} generate new types of Widom lines, are natural directions for future work.

By shifting the focus from conventional phase boundaries to the internal phase architecture, we show that degree heterogeneity generates crossover phenomena that differentiate states within an apparently single phase. Just as the Widom line deepened our understanding of fluids such as water and supercritical fluids, its network analog advances our understanding of complex systems by revealing hidden phase structures. Recognizing this mesoscopic architecture opens new directions for predicting, interpreting, and controlling collective behaviors in real-world networks.

\section*{Data Availability}
The data that support the findings of this study are available 
upon request from the corresponding author~\cite{data}.

\begin{acknowledgments}
BK was supported by the National Research Foundation of Korea with Grant No. RS-2023-00279802 and KENTECH Research Grant No. KRG-2021-01-007.
\end{acknowledgments}

\providecommand{\noopsort}[1]{}\providecommand{\singleletter}[1]{#1}%

\clearpage
\newpage

\appendix

\onecolumngrid



\title{\large Supplemental Material for\\
``Heterogeneous Network Topology Induces the Widom Line"}

\makeatletter
\renewcommand{\thesection}{S\arabic{section}}
\renewcommand{\theequation}{S\arabic{equation}}
\renewcommand{\thefigure}{S\arabic{figure}}
\renewcommand{\thetable}{S\arabic{table}}
\renewcommand{\bibnumfmt}[1]{[S#1]}

\setcounter{section}{0}
\setcounter{equation}{0}
\setcounter{figure}{0}
\setcounter{table}{0}
\setcounter{page}{1}
\makeatother

\begin{center}
\textbf{\large Supplemental Material for \\ ``Heterogeneous Network Topology Induces the Widom Line"}
\end{center}

\section{Phase Diagrams}

This section presents the detailed structure of ordered phases in the Ashkin-Teller (AT) and Invisible Potts (IP) models, respectively, illustrating how the Baxter and Ferro phases subdivide into distinct ordering regimes on scale-free networks. Both models display three characteristic states---distributed (Dis), hub-dominant (Hub), and supercritical (SC) orderings---separated by crossover temperatures that depend on degree heterogeneity and model parameters.

Monte Carlo simulations confirm the mean-field theoretical predictions for the order parameters. At the same time, degree-resolved magnetization analysis highlights the essential difference between distributed states, in which magnetization is more evenly distributed across different degree classes, and hub-dominant states, in which magnetic ordering predominantly localizes on high-degree nodes. These subdivisions of phases represent a universal feature of spin models on scale-free networks, where the interplay between thermal fluctuations and network heterogeneity gives rise to a rich phenomenology of ordering beyond that of conventional homogeneous systems.

\begin{figure}[!h]
\centering
\includegraphics[width=1.0\linewidth]{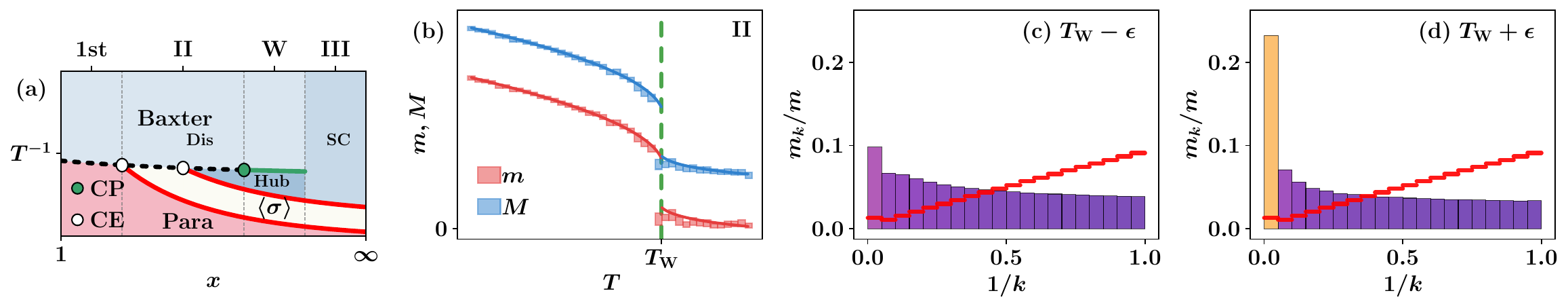}
\caption{(a) Schematic phase diagram based on mean-field analysis, showing the subdivision of the Baxter phase for the $x>1$ regime of the AT model.
\textbf{Dis}, \textbf{Hub}, and \textbf{SC} denote distributed, hub-dominant, and supercritical orderings, respectively.
(b) Monte Carlo results for the order parameters $m$ and $M$ versus temperature in the II regime of (a) with $x=1.23$ and $\lambda=3.9$; colored boxes indicate the mean $\pm$ one standard deviation, and solid lines indicate mean-field predictions. Monte Carlo simulations are performed on the annealed network Hamiltonian $\mathcal{H}_{\rm ann}$ using the Metropolis algorithm.
(c,d) Degree-resolved magnetization $m_k$ versus $1/k$ for (c) distributed and (d) hub-dominant states.
The interval $1/k \in [0,1]$ is divided into 20 equal bins; for each bin we compute $m_k = \sum_{i \in k} m_i k_i / (N \langle k \rangle)$, with $\sum_k m_k = m$.
Each bar shows the contribution of one bin to the total magnetization, while red staircases indicate the contributions of degree to gauge the expected spin contributions.
$T_{\rm w}$ $\approx 2.4674$ denotes the crossover temperature and $\epsilon$ a small parameter.}
\label{fig:figS5}
\end{figure}

\begin{figure}[!h]
\centering
\includegraphics[width=1.0\linewidth]{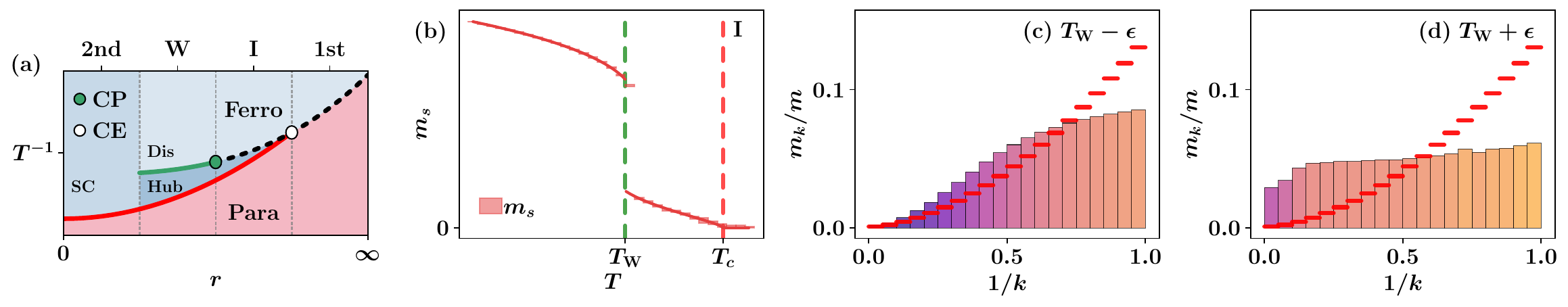}
\caption{(a) Schematic phase diagram based on mean-field analysis, showing the subdivision of the Ferro phase for the Invisible Potts model.
\textbf{Dis}, \textbf{Hub}, and \textbf{SC} denote distributed, hub-dominant, and supercritical orderings, respectively.
(b) Monte Carlo results for the order parameter $m$ versus temperature in the I regime of (a) with $r=8.4$ and $\lambda=4.8$; colored boxes indicate the mean $\pm$ one standard deviation, and solid lines indicate mean-field predictions. Monte Carlo simulations are performed on the annealed network Hamiltonian $\mathcal{H}_{\rm ann}$ using the Metropolis algorithm.
(c,d) Degree-resolved magnetization $m_k$ versus $1/k$ for (c) distributed and (d) hub-dominant states.
The interval $1/k \in [0,1]$ is divided into 20 equal bins; for each bin we compute $m_k = \sum_{i \in k} m_i k_i / (N \langle k \rangle)$, with $\sum_k m_k = m$.
Each bar shows the contribution of one bin to the total magnetization, while red staircases indicate the contributions of degree to gauge the expected spin contributions.
$T_{\rm w}$ $\approx 0.6365$ denotes the crossover temperature and $\epsilon$ a small parameter.}
\label{fig:figS6}
\end{figure}

\clearpage
\newpage

\section{Diverse Types of Phase Transitions}

\begin{figure}[!h]
\centering
\includegraphics[width=1.0\linewidth]{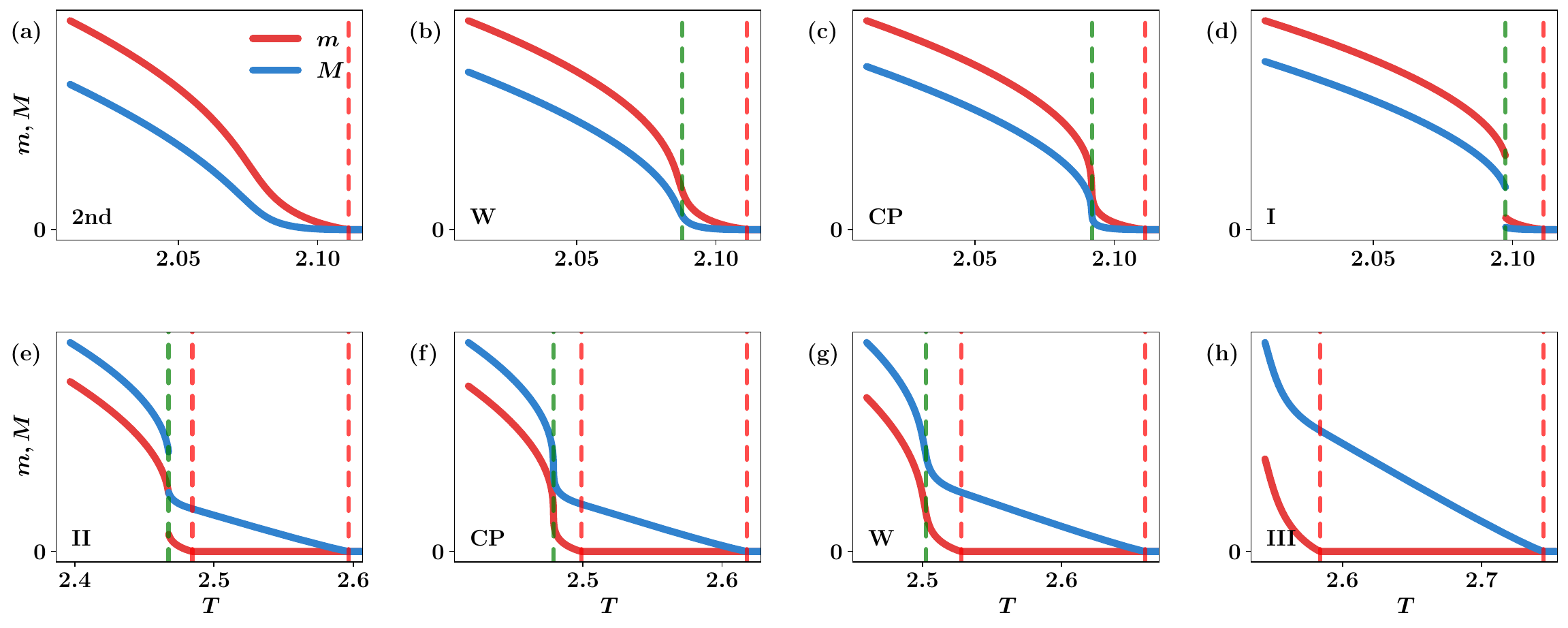}
\caption{Diverse Types of Phase Transition for Ashkin-Teller model. 
Panels (a-h) show the behavior of the order parameter $m$ and $M$.}
\label{fig:figS1}
\end{figure}

\begin{figure}[!h]
\centering
\includegraphics[width=1.0\linewidth]{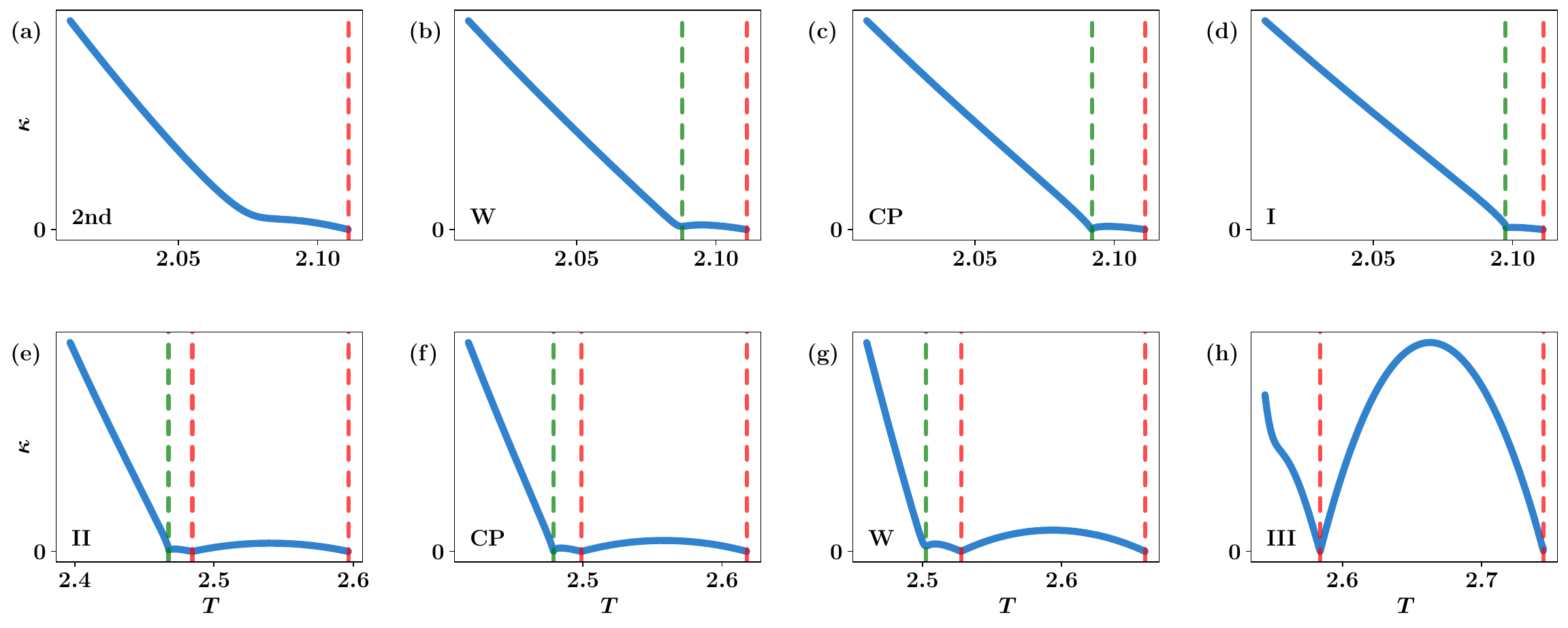}
\caption{Diverse Types of Phase Transition for Ashkin-Teller model. 
Panels (a-h) show the curvature $\kappa$ of the free energy.  
Continuous transitions occur when $\kappa$ vanishes, whereas crossovers emerge when $\kappa$ exhibits local minima without reaching zero.}
\label{fig:figS2}
\end{figure}

\begin{figure}[!h]
\centering
\includegraphics[width=1.0\linewidth]{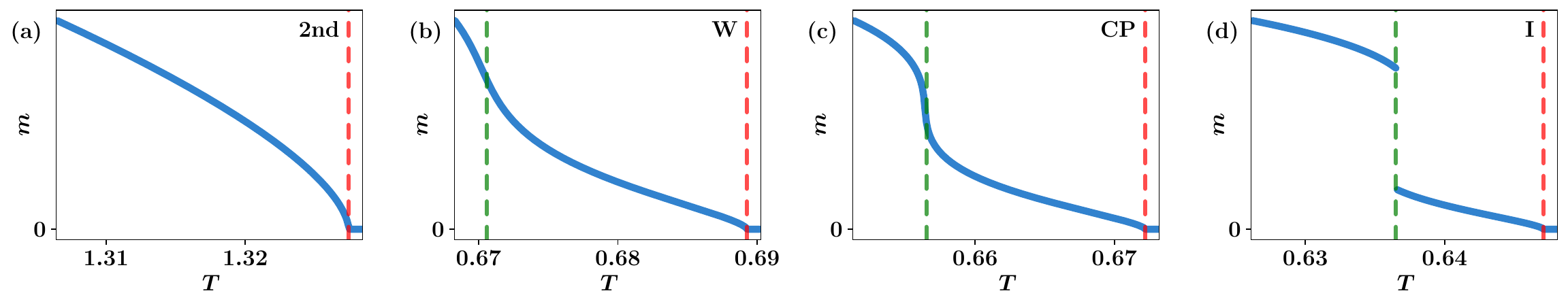}
\caption{Diverse types of phase transitions for the Invisible Potts model.  
Panels (a-d) show the behavior of the order parameter $m$.}
\label{fig:figS3}
\end{figure}

\begin{figure}[!h]
\centering
\includegraphics[width=1.0\linewidth]{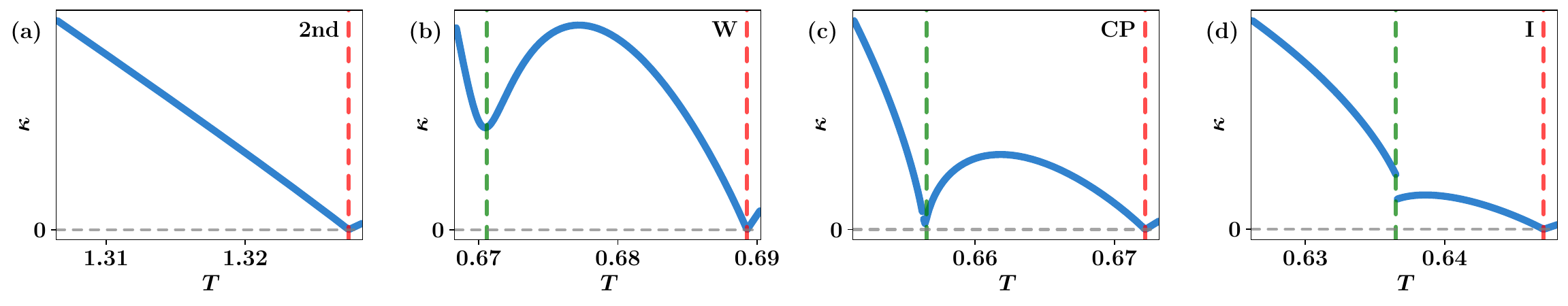}
\caption{Diverse types of phase transitions for the Invisible Potts model. 
Panels (a-d) show the curvature $\kappa$ of the free energy.  
Continuous transitions occur when $\kappa$ vanishes, whereas crossovers emerge when $\kappa$ exhibits local minima without reaching zero.}
\label{fig:figS4}
\end{figure}

This section illustrates the diverse types of phase transitions realized in the Ashkin-Teller (AT) and Invisible Potts (IP) models on scale-free networks.
Figs.~\ref{fig:figS1}--\ref{fig:figS4} demonstrate how different regimes give rise to discontinuous, continuous, and crossover transitions, characterized either by the order parameter $m$ or by the free-energy curvature $\kappa$.
Together, these results highlight the rich phenomenology of ordering transitions driven by degree heterogeneity in scale-free networks.

\subsection{The Ashkin-Teller model}

\subsection{The Invisible Potts model}

\clearpage
\newpage

\section{The Ashkin-Teller Model on Scale-Free Networks}

\subsection{Model Definition}

We analyze the Ashkin-Teller (AT) model on scale-free networks under the annealed network approximation~\cite{bianconi2002mean,jang2015ashkin,kim2021link}, in which the adjacency matrix is replaced by its degree-sequence average $\mathcal{A}_{ij} \rightarrow k_ik_j/(N\langle k\rangle)$. The Hamiltonian then becomes
\begin{align}
-\beta\mathcal{H}_{\rm ann} = \frac{K_2}{2 N\langle k\rangle}\sum_{i,j} k_ik_j \left(s_is_j + \sigma_i\sigma_j\right) \nonumber + \frac{K_4}{2 N\langle k\rangle}\sum_{i,j} k_ik_j s_i\sigma_i s_j\sigma_j,
\end{align}
where $K_2 = 1/T$ and $K_4 = x/T$. Each node $i$ hosts two Ising spins, $s_i$ and $\sigma_i$ ($= \pm 1$).

\subsection{Mean-Field Approximation}

Under $\mathcal{H}_{\rm ann}$, each spin interacts with all others weighted by degree, so defining the local mean fields $m_s^i = \langle s_i \rangle$, $m_\sigma^i = \langle \sigma_i \rangle$, and $m_{s\sigma}^i = \langle s_i \sigma_i \rangle$, the mean-field Hamiltonian becomes:
\begin{align}
-\beta\mathcal{H}_{\text{mf}} &= -\dfrac{1}{T}\sum_{\langle i,j \rangle}(m_s^i m_s^j + m_\sigma^i m_\sigma^j) - \dfrac{x}{T}\sum_{\langle i,j \rangle} m_{s\sigma}^i m_{s\sigma}^j + \dfrac{2}{T}\sum_{\langle i,j \rangle}(m_s^j s_i + m_\sigma^j \sigma_i) + \dfrac{2x}{T}\sum_{\langle i,j \rangle} m_{s\sigma}^j s_i \sigma_i.
\end{align}

\subsection{Free Energy Functional}

Assuming spin symmetry ($s_i \leftrightarrow \sigma_i$), we set $m_s = m_\sigma \equiv m$ and define $M = \langle s_i \sigma_i \rangle$. The free energy per node becomes:
\begin{align}
f &= \dfrac{1}{T} m^2 \langle k\rangle + \dfrac{1}{2}\dfrac{x}{T} M^2 \langle k\rangle - 2\int_{k_{\min}}^{\infty} \log\left[\cosh\left(\dfrac{m k}{T}\right)\right] P_d(k) \, dk - \int_{k_{\min}}^{\infty} \log\left[\cosh\left(\dfrac{x M k}{T}\right)\right] P_d(k) \, dk - \mathcal{B},
\end{align}
where
\begin{equation}
\mathcal{B} = \int_{k_{\min}}^{\infty} \log\left[1 + \tanh^2\left(\dfrac{m k}{T}\right) \tanh\left(\dfrac{x M k}{T}\right)\right] P_d(k) \, dk.
\end{equation}

\subsection{Self-Consistency Equations and Phase Structure}

Minimizing the free energy yields:
\begin{align}
m\langle k \rangle &= \int_{k_{\min}}^{\infty} \dfrac{\tanh({m k}/{T})\left[1+\tanh({x M k}/{T})\right]}{1+\tanh^2({m k}/{T})\tanh({x M k}/{T})} \, k P_d(k) \, dk, \\
M\langle k \rangle &= \int_{k_{\min}}^{\infty} \dfrac{\tanh({x M k}/{T})+\tanh^2({m k}/{T})}{1+\tanh^2({m k}/{T})\tanh({x M k}/{T})} \, k P_d(k) \, dk.
\end{align}

The resulting solutions define three phases:
\begin{itemize}
\item \textbf{Paramagnetic phase:} $m = M = 0$
\item \textbf{Baxter phase:} $m > 0$, $M > 0$
\item \textbf{$\langle \sigma s\rangle$ phase:} $m = 0$, $M > 0$
\end{itemize}

The case $m > 0$, $M = 0$ is forbidden due to coupling constraints.

\clearpage
\newpage

\subsection{Ginzburg-Landau Free Energy (For $3<\lambda<4$ case)}

The dominant contributions to the free energy differ between the weak-coupling ($x<1$) and strong-coupling ($x>1$) regimes. Accordingly, we expand the Landau free energy separately in these limits:

\begin{align}
f &\simeq 
\begin{cases}
\displaystyle
\dfrac{1}{T} m^2 \langle k\rangle + \dfrac{1}{2}\dfrac{x}{T} M^2 \langle k\rangle 
- 2\int_{k_{\min}}^{\infty} \log\left[\cosh\left(\dfrac{m k}{T}\right)\right] P_d(k) \, dk 
- \int_{k_{\min}}^{\infty} \log\left[\cosh\left(\dfrac{x M k}{T}\right)\right] P_d(k) \, dk 
- \mathcal{B}_{<} , 
& x < 1, \\[10pt]
\displaystyle
\dfrac{1}{T} m^2 \langle k\rangle + \dfrac{1}{2}\dfrac{x}{T} M^2 \langle k\rangle 
- 2\int_{k_{\min}}^{\infty} \log\left[\cosh\left(\dfrac{m k}{T}\right)\right] P_d(k) \, dk 
- \int_{k_{\min}}^{\infty} \log\left[\cosh\left(\dfrac{x M k}{T}\right)\right] P_d(k) \, dk
- \mathcal{B}_{>}, 
& x > 1,
\end{cases}
\end{align}
where
\begin{align}
    \mathcal{B}_{<} = \int_{k_{\min}}^{\infty} \left[\tanh^2\left(\frac{mk}{T}\right) \frac{xMk}{T}\right] P_d(k) \, dk, \quad \mathcal{B}_{>} = \int_{k_{\min}}^{\infty} \left[\tanh\left(\frac{x M k}{T}\right) \left(\frac{m k}{T}\right)^2\right] P_d(k) \, dk.
\end{align}

To reduce the Landau free energy to a single-variable form, we eliminate $M$ by expressing it in terms of $m$ via the self-consistency relation.

\begin{align}
    M \langle k \rangle \left(1 - \frac{x}{T} \cdot \frac{\langle k^2 \rangle}{\langle k \rangle} \right)
    \left\{
        \begin{array}{ll}
        \displaystyle
        \simeq - (\lambda-1) C(\lambda) \left(\frac{x M}{T}\right)^{\lambda-2} + D_{<}(\lambda) \left(\frac{m}{T}\right)^{\lambda-2} + \mathcal{O}(m^{\lambda-2}), 
        & \text{for } x < 1, \\[12pt]
        \displaystyle
        \simeq - (\lambda-1) C(\lambda) \left(\frac{x M}{T}\right)^{\lambda-2} + (\lambda-3)D_{>}(\lambda)\left(\frac{x M}{T}\right)^{\lambda-4} \left(\frac{m}{T}\right)^{2} & \text{for } x > 1, \\
        \displaystyle
        \to M_{*}\langle k \rangle \left(1 - \frac{x}{T} \cdot \frac{\langle k^2 \rangle}{\langle k \rangle} \right)
        + D_{>}(\lambda)\left(\frac{x M_{*}}{T}\right)^{\lambda-4} \left(\frac{m}{T}\right)^{2} + \mathcal{O}(m^{2}), 
        \end{array}
    \right.
\end{align}

Here, $M_{*}$ denotes the solution of the equation
\[
M \langle k \rangle \left(1 - \frac{x}{T} \cdot \frac{\langle k^2 \rangle}{\langle k \rangle} \right) = \int_{k_\min}^{\infty} k P_d(k) \tanh\left( \frac{xM}{T} \right) dk,
\]
which corresponds to the magnetization of a single-layer Ising model with effective coupling $x$.

Substituting this relation back into the expression for $f$ eliminates $M$ and produces a closed-form Landau free energy in terms of $m$. This allows us to extract the expansion coefficients up to the fourth order and to analyze the nature of the phase transition.

\paragraph{Weak Regime \texorpdfstring{$x<1$}{x<1}: }

\begin{align}
    f
    & \simeq 
    \dfrac{\langle k \rangle}{T}m^2 - 2 \displaystyle \int_{k_\min}^{\infty} P_d(k) \, dk \log \left[ \cosh(\dfrac{m k}{T}) \right]
    \cr
    & \quad 
    + \dfrac{\langle k \rangle}{ 2 T} x M^2 
    - \displaystyle \int_{k_\min}^{\infty} P_d(k) \, dk \log \left[ \cosh(\dfrac{x M k}{T}) \right] 
    - \displaystyle \int_{k_\min}^{\infty} P_d(k) \, dk \tanh^{2}\left(\dfrac{m k}{T}\right) \dfrac{xMk}{T} + \mathcal{O}(m^{4})
    \cr
    & \simeq 
    \dfrac{\langle k \rangle}{T}m^2 
    - 2 C(\lambda) \left(\dfrac{m}{T}\right)^{\lambda-1} 
    - \dfrac{x}{2 T} \left(\langle k \rangle - \langle k^2 \rangle/T \right)M^2 
    \cr
    & \simeq 
    \left(\dfrac{\langle k \rangle}{T} 
    - \dfrac{\langle k^{2} \rangle}{T^{2}}\right)m^2 
    - 2 C(\lambda) \left(\dfrac{m}{T}\right)^{\lambda-1} 
    - \dfrac{x\left[D_{<}(\lambda)\right]^{2}}{2T\left(\langle k \rangle - x\langle k^{2} \rangle/T\right)} \left(\dfrac{m}{T}\right)^{2(\lambda-2)},
\end{align}
where the coefficients are given by:
\begin{align}
    C(\lambda) = - \displaystyle \int_{0}^{\infty} P_d(\ell) \, d\ell \left\{ \log[2\cosh(\ell)] - \dfrac{1}{2} \ell^{2} \right\}, 
    \quad 
    D_{<}(\lambda) = \displaystyle \int_0^\infty \tanh^2(\ell) \ell^{1-\lambda} d\ell
\end{align}

\paragraph{Strong Regime \texorpdfstring{$x>1$}{x>1}: }
\begin{align}
    f
    & \simeq \dfrac{\langle k \rangle}{T}m^2 
    - 2 \displaystyle \int_{k_\min}^{\infty} P_d(k) \, dk \log \left[ \cosh(\dfrac{m k}{T}) \right] 
    \cr
    & \quad
    + \dfrac{\langle k \rangle}{ 2 T} x M^2 
    - \displaystyle \int_{k_\min}^{\infty} P_d(k) \, dk \log \left[ \cosh(\dfrac{x M k}{T}) \right] 
    - \displaystyle \int_{k_\min}^{\infty} P_d(k) \, dk \tanh\left(\dfrac{x M k}{T}\right) \left(\dfrac{m k}{T}\right)^2 + \mathcal{O}(m^{4})
    \cr
    & \simeq f(M_{*}) 
    + \left(\dfrac{\langle k \rangle}{T} - \dfrac{\langle k^{2} \rangle}{T^{2}} + C_{>}(\lambda)\right)m^2 
    - 2 C(\lambda) \left(\dfrac{m}{T}\right)^{\lambda-1} 
    - \dfrac{x}{2 T} \left(\langle k \rangle - \langle k^2 \rangle/T \right)\left(M-M_{*}\right)^2 + \mathcal{O}(m^{4}) \cr
    & \simeq f(M_{*}) 
    + \left(\dfrac{\langle k \rangle}{T} - \dfrac{\langle k^{2} \rangle}{T^{2}} + C_{>}(\lambda)\right)m^2 
    - 2 C(\lambda) \left(\dfrac{m}{T}\right)^{\lambda-1} 
    - \dfrac{x\left[D_{>}(\lambda)\right]^{2}}{2T\left(\langle k \rangle - x\langle k^{2} \rangle/T\right)} \left(\dfrac{x M_{*}}{T}\right)^{2(\lambda-4)} \left(\dfrac{m}{T}\right)^{4},
\end{align}
where the coefficients are given by:
\begin{align}
        C_{>}(\lambda) = \dfrac{1}{T^2}\displaystyle \int_{k_\min}^{\infty} k^2 P_d(k) \, dk \tanh\left(\dfrac{x M_{*} k}{T}\right), \quad D_{>}(\lambda) = \int_{0}^{\infty} \ell P_d(\ell) d\ell \tanh^{2}(\ell).
\end{align}
Here, the constant term $f(M_{*})$ corresponds to the Landau free energy of a single-layer Ising model with coupling strength $x$.

\subsection{Lower bound for Goldilocks zone}

The Goldilocks zone refers to the regime in which a Widom line appears. In this regime, sub-phases develop within the ordered phase, and transitions occur between them. For a Widom line to emerge, the Landau free energy must support at least two local minima; if only one exists, no subphase forms, and no Widom line appears. This condition requires that the distributed ordered state have an energy sufficiently lower than that of the hub-dominated state, with this energy difference ($|f_{\textrm{dis}} - f_{\textrm{hub}}|$) being maximized in the limit $x \to 1$. Using this property, we determine the lower bound $\lambda_d$ of the Goldilocks zone.

The numerical analysis shows that $3 < \lambda_d < 4$, so we restrict our attention to this regime. In the limit $x \to 1$, where $M \to m$, the Ginzburg-Landau free energy is reduced to
\begin{align}
    f(m) \simeq \dfrac{3}{2}\left(\dfrac{\langle k \rangle}{T} - \dfrac{\langle k^{2} \rangle}{T^{2}}\right)m^2
    - \left[3C(\lambda)+D(\lambda)\right]\left(\frac{m}{T}\right)^{\lambda-1}
    - \dfrac{\lambda-1}{4-\lambda}\left(k_{\min}\right)^{4-\lambda}\left(\frac{m}{T}\right)^{3},
\end{align}
where the coefficients are given by:
\begin{align}
        D(\lambda) = \displaystyle \int_{0}^{\infty} P_d(\ell) \, d\ell \ln\left[1 + \tanh^{3}(\ell)\right].
\end{align}

The mapping $M \to m$ equalizes the powers of the terms $C(\lambda)$ and $D(\lambda)$, which previously appeared at different orders. Although the $D_{<}(\lambda)$ and $D_{>}(\lambda)$ terms were higher in order than $C(\lambda)$ and therefore subdominant, their coefficients diverge as $x$ approaches unity, so that at $x=1$ all terms become comparable in order. 

To generate a distributed ordered state distinct from the hub-dominant phase, the $D(\lambda)$ term must be sufficiently large. This requires the coefficient of the $(\lambda-1)$-order term to be negative. If this coefficient is positive, the $D(\lambda)$ contribution is too weak, making the distributed and hub-dominant states thermodynamically indistinguishable. When the coefficient of $(\lambda-1)$ order ($\left[3C(\lambda)+D(\lambda)\right]$) changes sign from positive to negative, the spin-ordered states become distinguishable and a Widom line can emerge. Therefore, this point of sign change defines the boundary that determines the lower limit of the Goldilocks zone $\lambda_d \approx 3.503$.

\clearpage
\newpage

\subsection{Upper bound for Goldilocks zone}

\subsubsection{General}
For a system to exhibit Goldilocks zone behavior, the Ginzburg–Landau free energy must possess multiple local minima. 
For the Landau expansion $F = C_{2}m^{2} + C_{4}m^{4} + C_{6}m^{6} + \cdots$, the existence of multiple minima is governed by the coefficient signs. The $C_{2}$ term varies with temperature, while $C_{4}$ and $C_{6}$ depend on the control parameter. Although $C_{4}$ dictates the transition type, the emergence of additional minima near its sign change is based on $C_{6}$.

If $C_{6} > 0$ when $C_{4}$ changes sign, the free energy exhibits only the single minimum arising from $C_{2}$–$C_{4}$ competition. In contrast, if $C_{6} < 0$, extra minima appear through $C_{6}$–$C_{n>6}$ competition, producing sub-phases and enabling transitions among them.
Because $C_{6} > 0$ at the $C_{4}$ sign change rules out Widom-line formation, this condition defines the upper bound of the Goldilocks zone. In scale-free networks, this criterion is met for $\lambda > 7$, and therefore we focus our analysis on this regime.

\subsubsection{Weak Regime \texorpdfstring{$x<1$}{x<1}: }

The Landau free energy is given by
\begin{align}
f
&= \displaystyle
\dfrac{1}{T} m^2 \langle k\rangle + \dfrac{1}{2}\dfrac{x}{T} M^2 \langle k\rangle 
- 2\int_{k_{\min}}^{\infty} \log\left[\cosh\left(m k / T\right)\right] P_d(k) \, dk \cr
& \quad
- \int_{k_{\min}}^{\infty} \log\left[\cosh\left(x M k / T\right)\right] P_d(k) \, dk 
- \int_{k_{\min}}^{\infty} \log\left[1+\tanh^{2}\left(m k / T\right)\tanh\left(x M k / T\right)\right] P_d(k) \, dk, \cr
&\simeq \displaystyle
\dfrac{1}{T} m^2 \langle k\rangle + \dfrac{1}{2}\dfrac{x}{T} M^2 \langle k\rangle 
- 2\int_{k_{\min}}^{\infty} \log\left[\cosh\left(mk / T\right)\right] P_d(k) \, dk \cr
& \quad
- \int_{k_{\min}}^{\infty} \dfrac{1}{2} \left(\dfrac{xMk}{T}\right)^{2} P_d(k) \, dk
- \int_{k_{\min}}^{\infty} \tanh^{2}(mk/T) \dfrac{xMk}{T} P_d(k) \, dk, 
+ \int_{k_{\min}}^{\infty} \dfrac{1}{2} \tanh^{4}(mk/T) \left( \dfrac{x M k}{T} \right)^{2} P_d(k) \, dk, \cr
& \quad
+ \int_{k_{\min}}^{\infty} \dfrac{1}{3} \tanh^{2}(mk/T) \left(1 - \tanh^{4}\left(m k / T\right)\right) \left( \dfrac{xMk}{T} \right)^{3} P_d(k) \, dk.
\end{align}

To reduce the free energy to a single-variable function of $m$, we expand $M$ to powers of $m$:
\begin{align}
    M \langle k \rangle
    &= \int_{k_\min}^{\infty} k P_d(k) \, dk \, 
    \frac{\tanh^{2}\left(m k / T\right) + \tanh\left(x M k / T\right)}{1 + \tanh^{2}\left(m k / T\right)\tanh\left(x M k / T\right)} \cr
    &\simeq \int_{k_\min}^{\infty} k P_d(k) \, dk \left\{ 
    \tanh^{2}\left(m k / T\right)
    + \left(1 - \tanh^{4}\left(m k / T\right)\right) \frac{x M k}{T}
    + \tanh^{2}\left(m k / T\right) \left(1 - \tanh^{4}\left(m k / T\right)\right) \left(\frac{x  M k}{T}\right)^{2} 
    \right\}
    \cr
    &\simeq \int_{k_\min}^{\infty} k P_d(k) \, dk \left[ 
    \left(\frac{mk}{T} \right)^{2} 
    -\dfrac{2}{3} \left(\frac{mk}{T} \right)^{4}
    + \frac{x M k}{T}
    \right].
\end{align}

We obtain a self-consistency equation for $M$ to the second order in $(mk/T)$, denoted as $M_{2}$:
\begin{align}
    M_{2} 
    \left[ 
    \langle k \rangle - \int_{k_\min}^{\infty} 
    \frac{x k^{2}}{T} P_d(k) \, dk \, 
    \right]
    &= \int_{k_\min}^{\infty} k^{3} P_d(k) \, dk \, 
    \left(\frac{m}{T}\right)^{2}.
\end{align}

Thus,
\begin{align}
    M_{2} = 
    \frac{\displaystyle \int_{k_\min}^{\infty} k^{3} P_d(k) \, dk}
    {\langle k \rangle - \int_{k_\min}^{\infty} \frac{x k^{2}}{T} P_d(k) \, dk} 
    \left(\frac{m}{T}\right)^{2} 
    \equiv \dfrac{B(\lambda)}{A(\lambda)} 
    \left(\frac{m}{T}\right)^{2}.
\end{align}

Extending the expansion for $M$ to the fourth order in $(mk/T)$, denoted as $M_{4}$, yields
\begin{align}
    M_{4} 
    \left[ 
    \langle k \rangle - \int_{k_\min}^{\infty} \frac{x k^{2}}{T} P_d(k) \, dk \, 
    \right] = \int_{k_\min}^{\infty} k P_d(k) \, dk \, 
    \left[
    -\dfrac{2}{3}
    \right]
    \left(\frac{mk}{T} \right)^{4}.
  \end{align}

Finally, solving self-consistently for $M_4$ gives
\begin{align}
    M_{4} 
    \equiv \dfrac{C(\lambda)}{A(\lambda)} \left(\frac{m}{T}\right)^{4}, \,
    A(\lambda)
    = \left[ 
    \langle k \rangle - \int_{k_\min}^{\infty} 
    \frac{x k^{2}}{T} P_d(k) \, dk \, 
    \right], \, 
    C(\lambda)
    = \int_{k_\min}^{\infty} k^{5} P_d(k) \, dk \, 
    \left[ 
    -\dfrac{2}{3} 
    \right].
\end{align}

Using the self-consistency relation for $M = M_{2} + M_{4}$, we can expand the Landau free energy $f$ as a power series in $m$ up to sixth order:
\begin{align}
    f
    &\simeq \displaystyle
    \dfrac{1}{T} m^2 \langle k\rangle + \dfrac{1}{2}\dfrac{x}{T} M^2 \langle k\rangle 
    - 2\int_{k_{\min}}^{\infty} \log\left[\cosh\left(mk / T\right)\right] P_d(k) \, dk \cr
    & \quad
    - \int_{k_{\min}}^{\infty} \dfrac{1}{2} \left(\dfrac{xMk}{T}\right)^{2} P_d(k) \, dk
    - \int_{k_{\min}}^{\infty} \tanh^{2}(mk/T) \dfrac{xMk}{T} P_d(k) \, dk, \cr
    & \quad
    + \int_{k_{\min}}^{\infty} \dfrac{1}{2} \tanh^{4}(mk/T) \left( \dfrac{x M k}{T} \right)^{2} P_d(k) \, dk, \cr
    & \quad
    + \int_{k_{\min}}^{\infty} \dfrac{1}{3} \tanh^{2}(mk/T) \left(1 - \tanh^{4}\left(m k / T\right)\right) \left( \dfrac{xMk}{T} \right)^{3} P_d(k) \, dk,
\end{align}
\begin{align}
    f
    &\simeq \displaystyle
    \dfrac{1}{T} m^2 \langle k\rangle + \dfrac{1}{2}\dfrac{x}{T} M^2 \langle k\rangle 
    - 2\int_{k_{\min}}^{\infty} 
    \left[
    \dfrac{1}{2}\left(\dfrac{m k}{T}\right)^{2}
    - \dfrac{1}{12}\left(\dfrac{m k}{T}\right)^{4}
    + \dfrac{1}{45}\left(\dfrac{m k}{T}\right)^{6}
    \right] P_d(k) \, dk \cr
    & \quad
    - \int_{k_{\min}}^{\infty} \dfrac{1}{2}
    \left(\dfrac{xMk}{T}\right)^{2} P_d(k) \, dk 
    - \int_{k_{\min}}^{\infty} 
    \left( \dfrac{x M k}{T} \right)
    \left(\dfrac{m k}{T}\right)^{2}
    P_d(k) \, dk, 
    + \int_{k_{\min}}^{\infty} 
    \dfrac{2}{3} 
    \left( \dfrac{x M k}{T} \right) 
    \left(\dfrac{m k}{T}\right)^{4} 
    P_d(k) \, dk, 
\end{align}
\begin{align}
    f
    & \simeq f(M_c) + C_{2}(\lambda) \left(\dfrac{m}{T} \right)^{2} + C_{4}(\lambda) \left(\dfrac{m}{T} \right)^{4} + C_{6}(\lambda) \left(\dfrac{m}{T} \right)^{6} + \mathcal{O}(m^{6}).
\end{align}

where the coefficients are given by
\begin{align}
    C_{2}(\lambda)
    = T \langle k \rangle - \langle k^{2} \rangle,
    \,
    C_{4}(\lambda)
    = \dfrac{1}{6} \langle k^{4} \rangle 
    - \frac{x}{2T} \frac{B^{2}(\lambda)}{A(\lambda)},
    \,
    C_{6}(\lambda)
    = -\dfrac{2}{45} \langle k^{6} \rangle
    - \frac{x}{T} \frac{B(\lambda)C(\lambda)}{A(\lambda)} .
\end{align}

At the critical temperature where $C_{2}=0$, we identify the values of $(\lambda,T,x)$ satisfying $C_{4}=0$ and $C_{6}>0$. With three unknowns $(\lambda,T,x)$ and three equations, the system can be solved numerically, giving $\lambda_{u}\simeq 7.16$. This value coincides with the numerically determined boundary of the Goldilocks zone.

\clearpage
\newpage

\subsubsection{Strong Regime \texorpdfstring{$x>1$}{x>1}: }
The Landau free energy is given by
\begin{align}
    f
    &= \displaystyle
    \dfrac{1}{T} m^2 \langle k\rangle + \dfrac{1}{2}\dfrac{x}{T} M^2 \langle k\rangle 
    - 2\int_{k_{\min}}^{\infty} \log\left[\cosh\left(m k / T\right)\right] P_d(k) \, dk \cr
    & \quad
    - \int_{k_{\min}}^{\infty} \log\left[\cosh\left(x M k / T\right)\right] P_d(k) \, dk 
    - \int_{k_{\min}}^{\infty} \log\left[1+\tanh^{2}\left(m k / T\right)\tanh\left(x M k / T\right)\right] P_d(k) \, dk, \cr
    &= \displaystyle
    \dfrac{1}{T} m^2 \langle k\rangle + \dfrac{1}{2}\dfrac{x}{T} (M_{c} + \delta M)^2 \langle k\rangle 
    - 2\int_{k_{\min}}^{\infty} \log\left[\cosh\left(m k / T\right)\right] P_d(k) \, dk \\
    & \quad
    - \int_{k_{\min}}^{\infty} \log\left[\cosh\left(x (M_{c} + \delta M) k / T\right)\right] P_d(k) \, dk 
    - \int_{k_{\min}}^{\infty} \log\left[1+\tanh^{2}\left(m k / T\right)\tanh\left(x (M_{c} + \delta M) k / T\right)\right] P_d(k) \, dk \nonumber
\end{align}
\begin{align}
    &\simeq \displaystyle
    \dfrac{1}{T} m^2 \langle k\rangle + \dfrac{1}{2}\dfrac{x}{T} (M_{c} + \delta M)^2 \langle k\rangle 
    - 2\int_{k_{\min}}^{\infty} \log\left[\cosh\left(m k / T\right)\right] P_d(k) \, dk \cr
    & \quad
    - \int_{k_{\min}}^{\infty} \log\left[\cosh\left(x M_{c} k / T\right)\right] P_d(k) \, dk
    - \int_{k_{\min}}^{\infty} \log\left[1+\tanh^{2}\left(m k / T\right)\tanh\left(x M_{c} k / T\right)\right] P_d(k) \, dk, \cr
    & \quad
    - \int_{k_{\min}}^{\infty} \tanh\left(x M_{c} k / T\right) \left( \dfrac{x \delta M k}{T} \right) P_d(k) \, dk
    - \int_{k_{\min}}^{\infty} \dfrac{\tanh^{2}\left(m k / T\right) \left[1 - \tanh^{2}\left(x M_{c} k / T\right)\right]}{1 + \tanh^{2}\left(m k / T\right)\tanh\left(x M_{c} k / T\right)} \left( \dfrac{x \delta M k}{T} \right) P_d(k) \, dk, \cr
    & \quad
    - \int_{k_{\min}}^{\infty} \dfrac{1}{2} \left(1 - \tanh^{2}\left(x M_{c} k / T\right)\right) \left( \dfrac{x \delta M k}{T} \right)^{2} P_d(k) \, dk, \cr
    & \quad
    + \int_{k_{\min}}^{\infty} 
    \dfrac{\tanh^{4}\left(m k / T\right) \left(1 - \tanh^{4}\left(x M_{c} k / T\right)\right)}
    {2\left[1 + \tanh^{2}\left(m k / T\right)\tanh\left(x M_{c} k / T\right)\right]^{2}} 
    \left( \dfrac{x \delta M k}{T} \right)^{2} P_d(k) \, dk \cr
    & \quad
    + \int_{k_{\min}}^{\infty} 
    \dfrac{\tanh^{2}\left(m k / T\right) 2\tanh\left(x M_{c} k / T\right) \left(1 - \tanh^{2}\left(x M_{c} k / T\right)\right)}
    {2\left[1 + \tanh^{2}\left(m k / T\right)\tanh\left(x M_{c} k / T\right)\right]^{2}} 
    \left( \dfrac{x \delta M k}{T} \right)^{2} P_d(k) \, dk \cr
    & \quad
    + \int_{k_{\min}}^{\infty} \dfrac{1}{3} \tanh\left(x M_{c} k / T\right) \left(1 - \tanh^{2}\left(x M_{c} k / T\right)\right) \left( \dfrac{x \delta M k}{T} \right)^{3} P_d(k) \, dk,
\end{align}
where $M_c$ denotes the value of $M$ at the critical temperature $T_c$.

To reduce the free energy to a single-variable function of $m$, we expand $\delta M$ to powers of $m$:
\begin{align}
    M \langle k \rangle
    &= M_c \langle k \rangle + \delta M \langle k \rangle 
    = \int_{k_\min}^{\infty} k P_d(k) \, dk \, 
    \frac{\tanh^{2}\left(m k / T\right) + \tanh\left(x M k / T\right)}{1 + \tanh^{2}\left(m k / T\right)\tanh\left(x M k / T\right)} \cr
    &\simeq \int_{k_\min}^{\infty} k P_d(k) \, dk \left\{ 
    \frac{\tanh^{2}\left(m k / T\right) + \tanh\left(x M_{c} k / T\right)}{1 + \tanh^{2}\left(m k / T\right)\tanh\left(x M_{c} k / T\right)}
    + \frac{\left(1 - \tanh^{4}\left(m k / T\right)^{2}\right) \left(1 - \tanh^{2}\left(x M_c k / T\right)^{2}\right)}{\left[1 + \tanh^{2}\left(m k / T\right)\tanh\left(x M_{c} k / T\right)\right]^{2}} \frac{x \delta M k}{T} 
    \right. \cr
    &\quad - \left. 
    \frac{\left(1 - \tanh^{4}\left(m k / T\right)^{2}\right) \left(1 - \tanh^{2}\left(x M_c k / T\right)^{2}\right) \left( \tanh^{2}\left(m k / T\right) 
    + \tanh\left(x M_{c} k / T\right) \right)}{\left[1 + \tanh^{2}\left(m k / T\right)\tanh\left(x M_{c} k / T\right)\right]^{3}} \left(\frac{x \delta M k}{T}\right)^{2} 
    \right\}
\end{align}
\begin{align}
    M \langle k \rangle
    &\simeq \int_{k_\min}^{\infty} k P_d(k) \, dk \left[ 
    \tanh\left(x M_{c} k / T\right)
    + \left(1 - \tanh^{2}\left(x M_{c} k / T\right) \right) \left(\frac{mk}{T} \right)^{2} 
    \right] \cr
    & \quad + \int_{k_\min}^{\infty} k P_d(k) \, dk
    \left[ 
    -\dfrac{1}{3} \left(1 - \tanh^{2}\left(x M_{c} k / T\right) \right) \left(2 + 3 \tanh\left(x M_{c} k / T\right)\right)
    \right] 
    \left(\frac{mk}{T} \right)^{4} \cr
    &\quad + \int_{k_\min}^{\infty} k P_d(k) \, dk \left[ 
    \left(1 - \tanh^{2}\left(x M_{c} k / T\right) \right) \frac{x \delta M k}{T}
    - 2 \tanh\left(x M_{c} k / T\right) \left(1 - \tanh^{2}\left(x M_{c} k / T\right) \right) \frac{x \delta M k}{T}
    \left(\frac{mk}{T} \right)^{2} 
    \right] \cr
    &\quad + \int_{k_\min}^{\infty} k P_d(k) \, dk \left[ 
    -\tanh\left(x M_{c} k / T\right) \left(1 - \tanh^{2}\left(x M_{c} k / T\right) \right) \left(\frac{x \delta M k}{T}\right)^{2} 
    \right].
\end{align}

From the relation,
\begin{align}
    M_c \langle k \rangle = \int_{k_\min}^{\infty} k P_d(k) \, dk \, \tanh\left(x M_{c} k / T\right),
\end{align}
we obtain a self-consistency equation for $\delta M$ to the second order in $(mk/T)$, denoted as $\delta M_{2}$:
\begin{align}
    \delta M_{2} 
    \left[ 
    \langle k \rangle - \int_{k_\min}^{\infty} 
    \frac{x k^{2}}{T} P_d(k) \, dk \, 
    \left(1 - \tanh^{2}\left(x M_{c} k / T\right) \right) 
    \right]
    &= \int_{k_\min}^{\infty} k^{3} P_d(k) \, dk \, 
    \left(1 - \tanh^{2}\left(x M_{c} k / T\right) \right) 
    \left(\frac{m}{T}\right)^{2}.
\end{align}

Thus,
\begin{align}
    \delta M_{2} = 
    \frac{\displaystyle \int_{k_\min}^{\infty} k^{3} P_d(k) \, dk \, 
    \left(1 - \tanh^{2}\left(x M_{c} k / T\right) \right)}
    {\langle k \rangle - \int_{k_\min}^{\infty} \frac{x k^{2}}{T} P_d(k) \, dk \, \left(1 - \tanh^{2}\left(x M_{c} k / T\right) \right)} 
    \left(\frac{m}{T}\right)^{2} 
    \equiv \dfrac{B(\lambda)}{A(\lambda)} 
    \left(\frac{m}{T}\right)^{2}.
\end{align}

Extending the expansion for $\delta M$ to the fourth order in $(mk/T)$, denoted as $\delta M_{4}$, yields
\begin{align}
    &\delta M_{4} 
    \left[ 
    \langle k \rangle - \int_{k_\min}^{\infty} \frac{x k^{2}}{T} P_d(k) \, dk \, 
    \left(1 - \tanh^{2}\left(x M_{c} k / T\right) \right) 
    \right] \cr
    &= \int_{k_\min}^{\infty} k P_d(k) \, dk \, 
    \left[ 
    -\dfrac{1}{3} \left(1 - \tanh^{2}\left(x M_{c} k / T\right) \right) \left(2 + 3 \tanh\left(x M_{c} k / T\right)\right)
    \right] 
    \left(\frac{mk}{T} \right)^{4} \cr
    &\quad 
    + \int_{k_\min}^{\infty} k P_d(k) \, dk 
    \left[
    - 2 \tanh\left(x M_{c} k / T\right) \left(1 - \tanh^{2}\left(x M_{c} k / T\right) \right)
    \right]
    \frac{x \delta M k}{T} \left(\frac{mk}{T} \right)^{2} \cr
    &\quad 
    + \int_{k_\min}^{\infty} k P_d(k) \, dk 
    \left[ 
    - \tanh\left(x M_{c} k / T\right) \left(1 - \tanh^{2}\left(x M_{c} k / T\right) \right)     \right]
    \left(\frac{x \delta M k}{T}\right)^{2} 
  \end{align}

Finally, solving self-consistently for $\delta M_{4}$ gives
\begin{align}
    \delta M_{4} 
    &\equiv \dfrac{C_{1}(\lambda)+C_{2}(\lambda)+C_{3}(\lambda)}{A(\lambda)} \left(\frac{m}{T}\right)^{4}, \cr
    A(\lambda)
    &= \langle k \rangle 
    - \int_{k_\min}^{\infty} \frac{x k^{2}}{T} P_d(k) \, dk \, 
    \left(1 - \tanh^{2}\left(x M_{c} k / T\right) \right) \cr
    C_{1}(\lambda)
    &= \int_{k_\min}^{\infty} k^{5} P_d(k) \, dk \, 
    \left[ 
    -\dfrac{1}{3} \left(1 - \tanh^{2}\left(x M_{c} k / T\right) \right) \left(2 + 3 \tanh\left(x M_{c} k / T\right)\right)
    \right] \cr
    C_{2}(\lambda)
    &= \int_{k_\min}^{\infty} \dfrac{x k^{4}}{T} P_d(k) \, dk 
    \left[
    - 2 \tanh\left(x M_{c} k / T\right) \left(1 - \tanh^{2}\left(x M_{c} k / T\right) \right)
    \right] \cr
    C_{3}(\lambda)
    &= \int_{k_\min}^{\infty} \dfrac{x^{2} k^{3}}{T^{2}} P_d(k) \, dk 
    \left[ 
    - \tanh\left(x M_{c} k / T\right) \left(1 - \tanh^{2}\left(x M_{c} k / T\right) \right)  
    \right]
\end{align}

Using the self-consistency relation for $\delta M = \delta M_{2} + \delta M_{4}$, we can expand the Landau free energy $f$ as a power series in $m$ up to sixth order:
\begin{align}
    f
    &\simeq \displaystyle
    \dfrac{1}{T} m^2 \langle k\rangle + \dfrac{1}{2}\dfrac{x}{T} (M_{c} + \delta M)^2 \langle k\rangle 
    - 2\int_{k_{\min}}^{\infty} \log\left[\cosh\left(m k / T\right)\right] P_d(k) \, dk \cr
    & \quad
    - \int_{k_{\min}}^{\infty} \log\left[\cosh\left(x M_{c} k / T\right)\right] P_d(k) \, dk
    - \int_{k_{\min}}^{\infty} \log\left[1+\tanh^{2}\left(m k / T\right)\tanh\left(x M_{c} k / T\right)\right] P_d(k) \, dk, \cr
    & \quad
    - \int_{k_{\min}}^{\infty} \tanh\left(x M_{c} k / T\right) \left( \dfrac{x \delta M k}{T} \right) P_d(k) \, dk
    - \int_{k_{\min}}^{\infty} \dfrac{\tanh^{2}\left(m k / T\right) \left[1 - \tanh^{2}\left(x M_{c} k / T\right)\right]}{1 + \tanh^{2}\left(m k / T\right)\tanh\left(x M_{c} k / T\right)} \left( \dfrac{x \delta M k}{T} \right) P_d(k) \, dk, \cr
    & \quad
    - \int_{k_{\min}}^{\infty} \dfrac{1}{2} \left(1 - \tanh^{2}\left(x M_{c} k / T\right)\right) \left( \dfrac{x \delta M k}{T} \right)^{2} P_d(k) \, dk, \cr
    & \quad
    + \int_{k_{\min}}^{\infty} 
    \dfrac{\tanh^{2}\left(m k / T\right) \tanh^{2}\left(m k / T\right) \left(1 - \tanh^{4}\left(x M_{c} k / T\right)\right)}
    {2\left[1 + \tanh^{2}\left(m k / T\right)\tanh\left(x M_{c} k / T\right)\right]^{2}} 
    \left( \dfrac{x \delta M k}{T} \right)^{2} P_d(k) \, dk \cr
    & \quad
    + \int_{k_{\min}}^{\infty} 
    \dfrac{\tanh^{2}\left(m k / T\right) 2\tanh\left(x M_{c} k / T\right) \left(1 - \tanh^{2}\left(x M_{c} k / T\right)\right)}
    {2\left[1 + \tanh^{2}\left(m k / T\right)\tanh\left(x M_{c} k / T\right)\right]^{2}} 
    \left( \dfrac{x \delta M k}{T} \right)^{2} P_d(k) \, dk \cr
    & \quad
    + \int_{k_{\min}}^{\infty} \dfrac{1}{3} \tanh\left(x M_{c} k / T\right) \left(1 - \tanh^{2}\left(x M_{c} k / T\right)\right) \left( \dfrac{x \delta M k}{T} \right)^{3} P_d(k) \, dk,
\end{align}
\begin{align}
    f
    &\simeq \displaystyle
    \dfrac{1}{T} m^2 \langle k\rangle + \dfrac{1}{2}\dfrac{x}{T} (M_{c} + \delta M)^2 \langle k\rangle 
    - 2\int_{k_{\min}}^{\infty} 
    \left[
    \dfrac{1}{2}\left(\dfrac{m k}{T}\right)^{2}
    - \dfrac{1}{12}\left(\dfrac{m k}{T}\right)^{4}
    + \dfrac{1}{45}\left(\dfrac{m k}{T}\right)^{6}
    \right] P_d(k) \, dk \cr
    & \quad
    - \int_{k_{\min}}^{\infty} 
    \left[
    \cosh\left(x M_{c} k / T\right)
    \right] P_d(k) \, dk \cr
    & \quad
    - \int_{k_{\min}}^{\infty} 
    \left[
    \tanh\left(x M_c k / T\right)\left(\dfrac{m k}{T}\right)^{2}
    - \dfrac{1}{6}\left(4\tanh\left(x M_c k / T\right)+3\tanh^{2}\left(x M_c k / T\right)\right)
    \left(\dfrac{m k}{T}\right)^{4}
    \right] P_d(k) \, dk, \cr
    & \quad
    - \int_{k_{\min}}^{\infty} 
    \left[    
    \dfrac{1}{45}
    \left(
    17\tanh\left(x M_c k / T\right) 
    + 30\tanh^{2}\left(x M_c k / T\right)
    + 15\tanh^{3}\left(x M_c k / T\right)
    \right)\left(\dfrac{m k}{T}\right)^{6}
    \right] P_d(k) \, dk, \cr
    & \quad
    - \int_{k_{\min}}^{\infty} \tanh\left(x M_{c} k / T\right) 
    \left( \dfrac{x \delta M k}{T} \right) P_d(k) \, dk 
    - \int_{k_{\min}}^{\infty} 
    \left(1 - \tanh^{2}\left(x M_c k / T\right)\right)
    \left( \dfrac{x \delta M k}{T} \right)
    \left(\dfrac{m k}{T}\right)^{2}
    P_d(k) \, dk, \cr
    & \quad
    + \int_{k_{\min}}^{\infty} 
    \left[ 
    \dfrac{1}{3} \left(1 - \tanh^{2}\left(x M_{c} k / T\right) \right) \left(2 + 3 \tanh\left(x M_{c} k / T\right)\right)
    \right]
    \left( \dfrac{x \delta M k}{T} \right) 
    \left(\dfrac{m k}{T}\right)^{4} 
    P_d(k) \, dk, \cr
    & \quad
    - \int_{k_{\min}}^{\infty} \dfrac{1}{2} \left(1 - \tanh^{2}\left(x M_{c} k / T\right)\right) \left( \dfrac{x \delta M k}{T} \right)^{2} P_d(k) \, dk, \cr
    & \quad
    + \int_{k_{\min}}^{\infty} 
    \tanh\left(x M_{c} k / T\right)\left(1 - \tanh^{2}\left(x M_{c} k / T\right)\right)
    \left( \dfrac{x \delta M k}{T} \right)^{2} 
    \left(\dfrac{m k}{T}\right)^{2}
    P_d(k) \, dk \cr
    & \quad
    + \int_{k_{\min}}^{\infty} \dfrac{1}{3} \tanh\left(x M_{c} k / T\right) \left(1 - \tanh^{2}\left(x M_{c} k / T\right)\right) \left( \dfrac{x \delta M k}{T} \right)^{3} P_d(k) \, dk,
\end{align}
\begin{align}
    f
    & \simeq f(M_c) + C_{2}(\lambda) \left(\dfrac{m}{T} \right)^{2} + C_{4}(\lambda) \left(\dfrac{m}{T} \right)^{4} + C_{6}(\lambda) \left(\dfrac{m}{T} \right)^{6} + \mathcal{O}(m^{6}).
\end{align}

where the coefficients are given by
\begin{align}
    C_{2}(\lambda)
    &= T \langle k \rangle - \langle k^{2} \rangle
    - \int_{k_\min}^{\infty} k^{2} P_d(k) \, dk \, 
    \tanh\left(x M_c k / T\right), \\[10pt]
    C_{4}(\lambda)
    &= \dfrac{1}{6} \langle k^{4} \rangle 
    + \dfrac{1}{6} \int_{k_\min}^{\infty} k^{4} P_d(k) \, dk \, 
    \left(4\tanh\left(x M_c k / T\right)+3\tanh^{2}\left(x M_c k / T\right)\right) 
    - \frac{x}{2T} \frac{B^{2}(\lambda)}{A(\lambda)}, \\[10pt]
    C_{6}(\lambda)
    &= -\dfrac{2}{45} \langle k^{6} \rangle
    -\int_{k_\min}^{\infty} k^{6} P_d(k) \, dk \, 
    \left[
    \dfrac{1}{45}
    \left(
    17\tanh\left(x M_c k / T\right) 
    + 30\tanh^{2}\left(x M_c k / T\right)
    + 15\tanh^{3}\left(x M_c k / T\right)
    \right)
    \right] \cr
    & \quad
    - \frac{x}{T} \frac{B(\lambda)C_{1}(\lambda)}{A(\lambda)} 
    - \frac{x}{2T} \frac{B^{2}(\lambda)C_{2}(\lambda)}{A^{2}(\lambda)} 
    - \frac{x}{3T} \frac{B^{3}(\lambda)C_{3}(\lambda)}{A^{3}(\lambda)}.
\end{align}

At the critical temperature where $C_{2}=0$, we identify the values of $(\lambda,T,x)$ satisfying $C_{4}=0$ and $C_{6}>0$. With four unknowns $(M_c, \lambda,T,x)$ and four equations, the system can be solved numerically, giving $\lambda_{u}\simeq 7.21$. This value coincides with the numerically determined boundary of the Goldilocks zone.

\clearpage
\newpage

\section{The Invisible Potts Model on Scale-Free Networks}

\subsection{Model Definition}

We analyze the invisible Potts model on scale-free networks under the annealed network approximation, in which the adjacency matrix is replaced by its degree-sequence average $\mathcal{A}_{ij} \rightarrow k_ik_j/(N\langle k\rangle)$. The Hamiltonian then becomes
\begin{equation}
-\beta\mathcal{H}_{\rm ann} = \frac{K}{2 N\langle k\rangle} \sum_{i,j} \sum_{\alpha=1}^{q} k_i k_j \delta_{s_i,\alpha} \delta_{s_j,\alpha},
\end{equation}
where $K = 1/T$ is the ferromagnetic coupling constant. Each node $i$ hosts a spin $s_i$ that takes one of $q$ visible states ($\alpha = 1, \dots, q$) or $r$ hidden states ($\alpha = q+1, \dots, q+r$). Only visible states interact through alignment; hidden states contribute solely to entropy. Hereafter, we focus on the Ising-like case with $q = 2$.

\subsection{Mean-Field Approximation}

We define the average occupation probability of each spin state as follows.
\begin{equation}
\langle \delta_{s_i, \alpha} \rangle =
\begin{cases}
\mu_i & \text{for } \alpha = 1, \\
\nu_i & \text{for } \alpha = 2, \\
\rho_i & \text{for } \alpha = 3, \dots, 2+r,
\end{cases}
\end{equation}
subject to the normalization condition $\mu_i + \nu_i + r\rho_i = 1$. The local mean fields are defined as
\begin{equation}
m_i = \mu_i - \nu_i, \quad m_{r,i} = r \rho_i.
\end{equation}
Under $\mathcal{H}_{\rm ann}$, each spin interacts with all others weighted by degree, so the mean-field Hamiltonian becomes
\begin{equation}
-\beta\mathcal{H}_{\mathrm{mf}} = -\frac{1}{T} \sum_{\langle i,j \rangle} (\mu_i \mu_j + \nu_i \nu_j) + \frac{2}{T} \sum_{\langle i,j \rangle} \sum_{\alpha=1}^{2}\left[ \mu_j \delta_{s_i,\alpha} + \nu_j \delta_{s_i,\alpha} \right].
\end{equation}

\subsection{Free Energy Functional}

Under the annealed approximation, the global order parameters are defined as degree-weighted averages,
\begin{equation}
m = \sum_{i} m_i k_i / (N \langle k \rangle), \quad m_r = \sum_{i} m_{r,i} k_i / (N \langle k \rangle).
\end{equation}
The mean-field free energy per node is then given by
\begin{equation}
f = \frac{\langle k \rangle}{2T} \left[ (1 - m_r)^2 + m^2 \right] - \log \mathcal{Z},
\end{equation}
where the effective partition function $\mathcal{Z}$ is
\begin{equation}
\mathcal{Z} = \int_{k_{\min}}^{\infty} \left\{ e^{(1 - m_r + m)k/T} + e^{(1 - m_r - m)k/T} + r \right\} P_d(k) \, dk.
\end{equation}

\subsection{Self-Consistency Equations and Phase Structure}

Minimizing the free energy yields the self-consistency equations:
\begin{align}
m \langle k \rangle &= \int_{k_{\min}}^{\infty} \frac{e^{mk/T} - e^{-mk/T}}{e^{mk/T} + e^{-mk/T} + r e^{-(1 - m_r)k/T}} \, k P_d(k) \, dk, \\
m_r \langle k \rangle &= \int_{k_{\min}}^{\infty} \frac{r e^{-(1 - m_r)k/T}}{e^{mk/T} + e^{-mk/T} + r e^{-(1 - m_r)k/T}} \, k P_d(k) \, dk.
\end{align}
The system exhibits two distinct phases:
\begin{itemize}
\item \textbf{Paramagnetic phase:} $m = 0$, $m_r > 0$ --- hidden states dominate; visible spins remain disordered.
\item \textbf{Ferromagnetic phase:} $m > 0$, $m_r > 0$ --- visible spins develop spontaneous magnetization, while hidden states are partially occupied.
\end{itemize}

\subsection{Critical Temperature}

Near the critical temperature $T_c$, the order parameter $m$ vanishes, while the hidden-state occupancy $\bar{m}_r$ remains finite. The self-consistency relation for $\bar{m}_r$ reduces to
\begin{align}
[1 - \bar{m}_r] \langle k \rangle 
= \int_{k_\min}^{\infty} k P_d(k) \, dk \, \dfrac{1}{1 + \left( r/2 \right) e^{-(1 - \bar{m}_r)k/T}}.
\end{align}

At the critical point, the second-order coefficient of the free energy determines the onset of the transition:
\begin{align}
\dfrac{\langle k \rangle}{2 T_c} 
= \dfrac{1}{T_c^2} \int_{k_\min}^{\infty} k^2 P_d(k) \, dk \, \dfrac{1}{1 + \left( r/2 \right) e^{-(1 - \bar{m}_r)k/T_c}}.
\end{align}

Introducing a compact notation $R \equiv 1 - \bar{m}_r$ and assuming $R_c / T_c = \eta$, the two equations become
\begin{align}
R_c &
= \dfrac{1}{\langle k \rangle} \int_{k_\min}^{\infty} k P_d(k) \, dk \, 
\dfrac{1}{1 + \left( r/2 \right) e^{-R_c k / T_c}}
= \dfrac{1}{\langle k \rangle} \int_{k_\min}^{\infty} k P_d(k) \, dk \, 
\dfrac{1}{1 + \left( r/2 \right) e^{-\eta k}}, \\
T_c &
= \dfrac{1}{\langle k \rangle} \int_{k_\min}^{\infty} k^{2} P_d(k) \, dk \, 
\dfrac{1}{1 + \left( r/2 \right) e^{-R_c k / T_c}}
= \dfrac{1}{\langle k \rangle} \int_{k_\min}^{\infty} k^2 P_d(k) \, dk \, 
\dfrac{1}{1 + \left( r/2 \right) e^{-\eta k}}.
\end{align}

Combining these, we arrive at a closed self-consistency equation for $\eta$:
\begin{align}
\eta = \dfrac
{\displaystyle \int_{k_\min}^{\infty} k P_d(k) \, dk \, 
\dfrac{1}{1 + (r/2)e^{-\eta k}}}
{\displaystyle \int_{k_\min}^{\infty} k^2 P_d(k) \, dk \, 
\dfrac{1}{1 + (r/2)e^{-\eta k}}}.
\end{align}

Given a well-defined degree distribution $P_d(k)$ and the number of hidden states $r$, this self-consistent relation admits numerical or analytical solutions depending on the form of $P_d(k)$.

\subsection{Ginzburg-Landau Free Energy (For $3<\lambda<4$ case)}

The Landau free energy is given by
\begin{align}
f
&= \frac{\langle k \rangle}{2 T}(m^2 + R^2) - \int_{k_\min}^{\infty} P_d(k) \, dk \, \log \left[ 2 e^{R k/T} \cosh(mk/T)+r 
\right] \cr
&= \frac{\langle k \rangle}{2 T}[m^2 + (R_c + \delta R)^2] 
- \int_{k_\min}^{\infty} P_d(k) \, dk \, \log \left[ 
2 e^{(R_c + \delta R) k/T} \cosh(mk/T)+r 
\right] \cr
&\simeq \frac{\langle k \rangle}{2 T}[m^2 + (R_c + \delta R)^2] 
- \int_{k_\min}^{\infty} P_d(k) \, dk \, 
\log \left[ 
2 e^{R_c k/T}\cosh(mk/T) + r 
\right] \cr
&\quad - \int_{k_\min}^{\infty} P_d(k) \, dk \left\{ 
\frac{2 e^{R_c k/T}\cosh(mk/T)}
{2 e^{R_c k/T}\cosh(mk/T) + r} \frac{\delta R k}{T}
+ \frac{ r e^{R_c k/T}\cosh(mk/T)}
{[2 e^{R_c k/T}\cosh(mk/T) + r]^{2}} \left(\frac{\delta R k}{T} \right)^{2} 
\right\}, 
\end{align}
where $R_c$ denotes the value of $R$ at the critical temperature $T_c$.

To reduce the free energy to a single-variable function of $m$, we expand $\delta R$ to powers of $m$:
\begin{align}
R \langle k \rangle
&= R_c \langle k \rangle + \delta R \langle k \rangle \cr
&= \int_{k_\min}^{\infty} k P_d(k) \, dk \, 
\frac{2 e^{R k/T}\cosh(mk/T)}{2 e^{R k/T}\cosh(mk/T)+r} \cr
&\simeq \int_{k_\min}^{\infty} k P_d(k) \, dk 
\left\{ 
\frac{2 e^{R_c k/T}\cosh(mk/T)}{2 e^{R_c k/T}\cosh(mk/T)+r}
+ \frac{2r e^{R_c k/T}\cosh(mk/T)}{[2 e^{R_c k/T}\cosh(mk/T) + r]^{3}} \frac{\delta R k}{T} 
\right\} \cr
&\simeq \int_{k_\min}^{\infty} k P_d(k) \, dk 
\left[ 
\frac{2 e^{R_c k/T}}{2 e^{R_c k/T}+r}
+ \frac{r e^{R_c k/T}}{(2 e^{R_c k/T} + r)^{2}} \left(\frac{mk}{T} \right)^{2}
+\frac{2r e^{R_c k/T}}{(2 e^{R_c k/T} + r)^{2}} 
\frac{\delta R k}{T}
\right].
\end{align}

From the relation,
\begin{align}
    R_c \langle k \rangle 
    = \int_{k_\min}^{\infty} k P_d(k) \, dk \, 
    \frac{2 e^{R_c k/T}}{2 e^{R_c k/T}+r},
\end{align}
we obtain a self-consistency equation for $\delta R$ to the second order in $(mk/T)$, denoted as $\delta R_{2}$:
\begin{align}
    \delta R_{2}
    \left[ \langle k \rangle - \int_{k_\min}^{\infty} 
    \frac{k^{2}}{T} P_d(k) \, dk \, \frac{2 re^{R_c k/T}}{\big(2 e^{R_c k/T} + r\big)^{2}} \right]
    &= \int_{k_\min}^{\infty} k^{3} P_d(k) \, dk \, 
    \frac{r e^{R_c k/T}}{\big(2 e^{R_c k/T} + r\big)^{2}} 
    \left(\frac{m}{T}\right)^{2}.
\end{align}

Thus,
\begin{align}
    \delta R_{2} = 
    \frac{\displaystyle \int_{k_\min}^{\infty} k^{3} P_d(k) \, dk \, \frac{r e^{R_c k/T}}{\big(2 e^{R_c k/T} + r\big)^{2}}}
    {\displaystyle \langle k \rangle - \int_{k_\min}^{\infty} \frac{k^{2}}{T} P_d(k) \, dk \, \frac{2r e^{R_c k/T}}{\big(2 e^{R_c k/T} + r\big)^{2}}} 
    \left(\frac{m}{T}\right)^{2} 
    \equiv \dfrac{B(\lambda)}{A(\lambda)} \left(\frac{m}{T}\right)^{2}.
\end{align}

Using the self-consistency relation for $\delta R$, we can expand the Landau free energy $f$ as a power series in $m$ up to the fourth order:
\begin{align}
    f
    & \simeq \frac{\langle k \rangle}{2 T}[m^2 + (R_c + \delta R)^2] 
    - \int_{k_\min}^{\infty} P_d(k) \, dk \, \log 
    \left[ 
    2 e^{R_c k/T}\cosh(mk/T) + r 
    \right] \cr
    &\quad - \int_{k_\min}^{\infty} P_d(k) \, dk 
    \left\{ 
    \frac{2 e^{R_c k/T}\cosh(mk/T)}{2 e^{R_c k/T}\cosh(mk/T) + r} 
    \frac{\delta R k}{T}
    + \frac{r e^{R_c k/T}\cosh(mk/T)}
    {[2 e^{R_c k/T}\cosh(mk/T) + r]^{2}} \left(\frac{\delta R k}{T} \right)^{2} \right\},
\end{align}
\begin{align}    
    f
    & \simeq \dfrac{\langle k \rangle}{2 T}[m^2 + (R_c + \delta R)^2] \cr
    & \quad - \int_{k_\min}^{\infty} P_d(k) \, dk \left\{
    \log(2e^{R_c k/T} + r)
    + \dfrac{e^{R_c k/T}}{2e^{R_c k/T} + r} 
    \left(\dfrac{mk}{T} \right)^{2} 
    \right\} \cr
    & \quad 
    - \int_{0}^{\infty} P_d(k) \, dk \left\{ 
    \log \left[ 
    2 e^{R_c k/T}\cosh(mk/T) + r \right]
    - \log(2e^{R_c k/T} + r)
    - \dfrac{e^{R_c k/T}}{2e^{R_c k/T} + r} 
    \left(\dfrac{mk}{T} \right)^{2} 
    \right\} \cr    
    & \quad 
    + \int_{0}^{k_\min} P_d(k) \, dk \left[ 
    - \dfrac{e^{R_c k/T}(4e^{R_c k/T}-r)}{12(2e^{R_c k/T} + r)^{2}} 
    \left(\dfrac{mk}{T} \right)^{4} 
    \right] \cr
    & \quad - \int_{k_\min}^{\infty} P_d(k) \, dk \, 
    \left\{
    \dfrac{2e^{R_c k/T}}{2e^{R_c k/T}+r} 
    \frac{\delta R k}{T} 
    + \dfrac{re^{R_c k/T}}{(2e^{R_c k/T}+r)^{2}} 
    \frac{\delta R k}{T} 
    \left(\dfrac{mk}{T} \right)^{2}
    - \dfrac{re^{R_c k/T}}{(2e^{R_c k/T}+r)^{2}} 
    \left(\frac{\delta R k}{T}\right)^{2}
    \right\}.
\end{align}
\begin{align}
    f
    & \simeq \dfrac{\langle k \rangle}{2 T} R_c^{2} 
    + C_{2}(\lambda) \left(\dfrac{m}{T} \right)^{2} 
    + C_{\lambda-1}(\lambda) \left(\dfrac{m}{T} \right)^{\lambda-1} 
    + C_{4}(\lambda) \left(\dfrac{m}{T} \right)^{4} 
    + \mathcal{O}(m^{4}).
\end{align}

where the coefficients are given by
\begin{align}
    C_{2}(\lambda)
    &= \dfrac{T \langle k \rangle}{2} 
    - \int_{k_\min}^{\infty} k^{2} P_d(k) \, dk \, \dfrac{e^{R_c k/T}}{2e^{R_c k/T} + r}, 
    \cr
    C_{\lambda}(\lambda) 
    &= - \int_{0}^{\infty} P_d(\ell) \, d\ell \left\{ \log[2\cosh(\ell)] - \tfrac{1}{2} \ell^{2} \right\}, \cr
    C_{4}(\lambda)
    &= \int_{0}^{k_\min} k^{4} P_d(k) \, dk \, \left[
    - \dfrac{e^{R_c k/T}(4e^{R_c k/T}-r)}{12(2e^{R_c k/T} + r)^{2}} 
    \right]
    - \frac{1}{2T} \frac{B^{2}(\lambda)}{A(\lambda)}.
\end{align}


\begin{figure}[!h]
\centering
\includegraphics[width=1.0\linewidth]{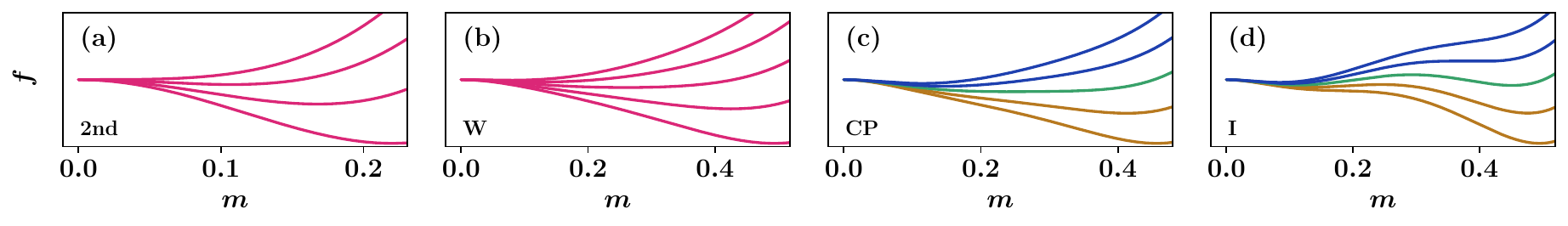}
\caption{GL free energy landscapes for the Invisible Potts model.  
Amber and navy curves represent entropy-driven alignment of visible states and hub-localized visible-state clusters, respectively.  
Green curves indicate transitions—either continuous or crossover—between these distinct ordered states.  
Crossovers emerge as smooth transitions between the two ordering modes.  
The deep pink curves denote a supercritical-like state where the two ordered states become thermodynamically indistinguishable.}
\label{fig:figS7}
\end{figure}

\subsection{Lower bound for Goldilocks zone}

The IP model follows the same general principle as the AT model, but with a simpler mechanism. Here, the energy of the distributed state is minimized in the limit $r \to \infty$. In this limit, the quartic term dominates throughout the range $3 < \lambda < 4$, generating a distributed ordered state distinct from the hub-dominant phase and thereby allowing for Widom-line formation. 
Numerical analysis verifies that the Widom lines and various phase transitions emerge only for $3 < \lambda < \lambda_{u}$, establishing the lower bound $\lambda_d = 3$ for the IP model.

\clearpage
\newpage

\subsection{Upper bound for Goldilocks zone}

As in the AT model, the absence of a Widom line in the IP model is determined by the analytic criterion $C_{6}>0$, where the sign of $C_{4}$ changes. This sets the upper bound of the Goldilocks zone. On scale-free networks, this condition holds for $\lambda>7$, and therefore we restrict our analysis to this regime.

The GL free energy is given by
\begin{align}
f
&= \frac{\langle k \rangle}{2 T}(m^2 + R^2) - \int_{k_\min}^{\infty} P_d(k) \, dk \, \log \left[ 2 e^{R k/T} \cosh(mk/T)+r 
\right] \cr
&= \frac{\langle k \rangle}{2 T}[m^2 + (R_c + \delta R)^2] 
- \int_{k_\min}^{\infty} P_d(k) \, dk \, \log \left[ 
2 e^{(R_c + \delta R) k/T} \cosh(mk/T)+r 
\right] \cr
&\simeq \frac{\langle k \rangle}{2 T}[m^2 + (R_c + \delta R)^2] 
- \int_{k_\min}^{\infty} P_d(k) \, dk \, 
\log \left[ 
2 e^{R_c k/T}\cosh(mk/T) + r 
\right] \cr
&\quad - \int_{k_\min}^{\infty} P_d(k) \, dk \left\{ 
\frac{2 e^{R_c k/T}\cosh(mk/T)}
{2 e^{R_c k/T}\cosh(mk/T) + r} \frac{\delta R k}{T}
+ \frac{ r e^{R_c k/T}\cosh(mk/T)}
{[2 e^{R_c k/T}\cosh(mk/T) + r]^{2}} \left(\frac{\delta R k}{T} \right)^{2} 
\right\} \cr
&\quad + \int_{k_\min}^{\infty} P_d(k) \, dk \left\{ 
\frac{r e^{R_c k/T}\cosh(mk/T)\left(2 e^{R_c k/T}\cosh(mk/T)-r\right)}
{3[2 e^{R_c k/T}\cosh(mk/T) + r]^{3}} 
\left(\frac{\delta R k}{T} \right)^{3} + \mathcal{O}\left[\left(\frac{\delta R k}{T} \right)^{3}\right]
\right\},
\end{align}
where $R_c$ denotes the value of $R$ at the critical temperature $T_c$.

To reduce the free energy to a single-variable function of $m$, we expand $\delta R$ to powers of $m$:
\begin{align}
R \langle k \rangle
&= R_c \langle k \rangle + \delta R \langle k \rangle \cr
&= \int_{k_\min}^{\infty} k P_d(k) \, dk \, 
\frac{2 e^{R k/T}\cosh(mk/T)}{2 e^{R k/T}\cosh(mk/T)+r} \cr
&\simeq \int_{k_\min}^{\infty} k P_d(k) \, dk 
\left\{ 
\frac{2 e^{R_c k/T}\cosh(mk/T)}{2 e^{R_c k/T}\cosh(mk/T)+r}
+ \frac{2r e^{R_c k/T}\cosh(mk/T)}{[2 e^{R_c k/T}\cosh(mk/T) + r]^{3}} \frac{\delta R k}{T} 
\right\} \cr
&\quad - \int_{k_\min}^{\infty} k P_d(k) \, dk 
\left\{ 
\frac{r e^{R_c k/T}\cosh(mk/T) \left[2 e^{R_c k/T}\cosh(mk/T)-r\right]}
{[2 e^{R_c k/T}\cosh(mk/T) + r]^{2}} \left(\frac{\delta R k}{T}\right)^{2} 
+ \mathcal{O}\left[\left(\frac{\delta R k}{T} \right)^{3}\right] 
\right\} \cr
&\simeq \int_{k_\min}^{\infty} k P_d(k) \, dk 
\left[ 
\frac{2 e^{R_c k/T}}{2 e^{R_c k/T}+r}
+ \frac{r e^{R_c k/T}}{(2 e^{R_c k/T} + r)^{2}} \left(\frac{mk}{T} \right)^{2}
- \frac{r e^{R_c k/T}(10 e^{R_c k/T}-r)}{12(2 e^{R_c k/T} + r)^{3}} \left(\frac{mk}{T} \right)^{4} 
\right]\cr
&\quad + \int_{k_\min}^{\infty} k P_d(k) \, dk 
\left[ 
\frac{2r e^{R_c k/T}}{(2 e^{R_c k/T} + r)^{2}} 
\frac{\delta R k}{T}
- \frac{r e^{R_c k/T}(2 e^{R_c k/T}-r)}{(2 e^{R_c k/T} + r)^{3}} 
\frac{\delta R k}{T} \left(\frac{mk}{T} \right)^{2} 
\right] \cr
&\quad - \int_{k_\min}^{\infty} k P_d(k) \, dk 
\left[ 
\frac{r e^{R_c k/T}(2 e^{R_c k/T}-r)}{(2 e^{R_c k/T} + r)^{3}} 
\left(\frac{\delta R k}{T}\right)^{2} 
\right].
\end{align}

From the relation,
\begin{align}
    R_c \langle k \rangle 
    = \int_{k_\min}^{\infty} k P_d(k) \, dk \, 
    \frac{2 e^{R_c k/T}}{2 e^{R_c k/T}+r},
\end{align}
we obtain a self-consistency equation for $\delta R$ to the second order in $(mk/T)$, denoted as $\delta R_{2}$:
\begin{align}
    \delta R_{2}
    \left[ \langle k \rangle - \int_{k_\min}^{\infty} 
    \frac{k^{2}}{T} P_d(k) \, dk \, \frac{2 re^{R_c k/T}}{\big(2 e^{R_c k/T} + r\big)^{2}} \right]
    &= \int_{k_\min}^{\infty} k^{3} P_d(k) \, dk \, 
    \frac{r e^{R_c k/T}}{\big(2 e^{R_c k/T} + r\big)^{2}} 
    \left(\frac{m}{T}\right)^{2}.
\end{align}

Thus,
\begin{align}
    \delta R_{2} = 
    \frac{\displaystyle \int_{k_\min}^{\infty} k^{3} P_d(k) \, dk \, \frac{r e^{R_c k/T}}{\big(2 e^{R_c k/T} + r\big)^{2}}}
    {\displaystyle \langle k \rangle - \int_{k_\min}^{\infty} \frac{k^{2}}{T} P_d(k) \, dk \, \frac{2r e^{R_c k/T}}{\big(2 e^{R_c k/T} + r\big)^{2}}} 
    \left(\frac{m}{T}\right)^{2} 
    \equiv \dfrac{B(\lambda)}{A(\lambda)} \left(\frac{m}{T}\right)^{2}.
\end{align}

Extending the expansion for $\delta R$ to the fourth order in $(mk/T)$, denoted as $\delta R_{4}$, yields
\begin{align}
    &\delta R_{4} 
    \left[ 
    \langle k \rangle - \int_{k_\min}^{\infty} \frac{k^{2}}{T} P_d(k) \, dk \, 
    \frac{2 e^{R_c k/T}}{\big(2 e^{R_c k/T} + r\big)^{2}} 
    \right] \cr
    &= - \int_{k_\min}^{\infty} k P_d(k) \, dk \, 
    \frac{r e^{R_c k/T}(10 e^{R_c k/T}-r)}{12(2 e^{R_c k/T} + r)^{3}} 
    \left(\frac{mk}{T} \right)^{4} \cr
    &\quad - \int_{k_\min}^{\infty} k P_d(k) \, dk 
    \left[ 
    \frac{r e^{R_c k/T}(2 e^{R_c k/T}-r)}{(2 e^{R_c k/T} + r)^{3}} 
    \frac{\delta R k}{T} \left(\frac{mk}{T} \right)^{2}
    + \frac{r e^{R_c k/T}(2 e^{R_c k/T}-r)}{(2 e^{R_c k/T} + r)^{3}} 
    \left(\frac{\delta R k}{T}\right)^{2} 
    \right].
\end{align}

Finally, solving self-consistently for $\delta R_{4}$ gives
\begin{align}
    \delta R_{4} 
    &\equiv \dfrac{C_{1}(\lambda)+C_{2}(\lambda)+C_{3}(\lambda)}{A(\lambda)} \left(\frac{m}{T}\right)^{4} \cr
    A(\lambda)
    & = \displaystyle \langle k \rangle - \int_{k_\min}^{\infty} \frac{k^{2}}{T} P_d(k) \, dk \, \frac{2 e^{R_c k/T}}{\big(2 e^{R_c k/T} + r\big)^{2}} \cr
    C_{1}(\lambda)
    & = \displaystyle - \int_{k_\min}^{\infty} k^{5} P_d(k) \, dk 
    \frac{r e^{R_c k/T}(10 e^{R_c k/T}-r)}{12(2 e^{R_c k/T} + r)^{3}}
    \cr
    C_{2}(\lambda)
    & = \displaystyle - \int_{k_\min}^{\infty} \frac{k^{4}}{T} P_d(k) \, dk 
    \delta R_2\frac{r e^{R_c k/T}(2 e^{R_c k/T}-r)}{(2 e^{R_c k/T} + r)^{3}}
    \cr
    C_{3}(\lambda)
    & = \displaystyle - \int_{k_\min}^{\infty} \frac{k^{3}}{T^{2}} P_d(k) \, dk 
    \delta R_2^{2}\frac{r e^{R_c k/T}(2 e^{R_c k/T}-r)}{(2 e^{R_c k/T} + r)^{3}}.
\end{align}

Using the self-consistency relation for $\delta R = \delta R_{2} + \delta R_{4}$, we can expand the Landau free energy $f$ as a power series in $m$ up to the sixth order:
\begin{align}
    f
    & \simeq \frac{\langle k \rangle}{2 T}[m^2 + (R_c + \delta R)^2] 
    - \int_{k_\min}^{\infty} P_d(k) \, dk \, \log 
    \left[ 
    2 e^{R_c k/T}\cosh(mk/T) + r 
    \right] \cr
    &\quad - \int_{k_\min}^{\infty} P_d(k) \, dk 
    \left\{ 
    \frac{2 e^{R_c k/T}\cosh(mk/T)}{2 e^{R_c k/T}\cosh(mk/T) + r} 
    \frac{\delta R k}{T}
    + \frac{r e^{R_c k/T}\cosh(mk/T)}
    {[2 e^{R_c k/T}\cosh(mk/T) + r]^{2}} \left(\frac{\delta R k}{T} \right)^{2} \right\} \cr
    &\quad + \int_{k_\min}^{\infty} P_d(k) \, dk 
    \left\{ 
    \frac{r e^{R_c k/T}\cosh(mk/T)\left(2 e^{R_c k/T}\cosh(mk/T)-r\right)}
    {3[2 e^{R_c k/T}\cosh(mk/T) + r]^{3}} 
    \left(\frac{\delta R k}{T} \right)^{3} 
    + \mathcal{O}\!\left[\left(\frac{\delta R k}{T} \right)^{3}\right]\right\}, \cr
    f
    & \simeq \dfrac{\langle k \rangle}{2 T}[m^2 + (R_c + \delta R)^2] \cr
    & \quad - \int_{k_\min}^{\infty} P_d(k) \, dk \log(2e^{R_c k/T} + r)\cr
    & \quad - \int_{k_\min}^{\infty} P_d(k) \, dk 
    \left\{ 
    \dfrac{e^{R_c k/T}}{2e^{R_c k/T} + r} 
    \left(\dfrac{mk}{T} \right)^{2} 
    - \dfrac{e^{R_c k/T}(4e^{R_c k/T}-r)}{12(2e^{R_c k/T} + r)^{2}} 
    \left(\dfrac{mk}{T} \right)^{4} 
    + \dfrac{e^{R_c k/T}(64e^{2R_c k/T}-26re^{R_c k/T}+r^{2})}{360(2e^{R_c k/T} + r)^{3}} 
    \left(\dfrac{mk}{T} \right)^{6}
    \right\} \cr
    & \quad - \int_{k_\min}^{\infty} P_d(k) \, dk \, 
    \left\{
    \dfrac{2e^{R_c k/T}}{2e^{R_c k/T}+r} 
    \frac{\delta R k}{T} 
    + \dfrac{re^{R_c k/T}}{(2e^{R_c k/T}+r)^{2}} 
    \frac{\delta R k}{T} 
    \left(\dfrac{mk}{T} \right)^{2}
    - \dfrac{re^{R_c k/T}(10e^{R_c k/T}-r)}{12(2e^{R_c k/T}+r)^{3}} 
    \frac{\delta R k}{T} 
    \left(\dfrac{mk}{T} \right)^{4} \right\} \cr
    & \quad - \int_{k_\min}^{\infty} P_d(k) \, dk \, 
    \left\{
    \dfrac{re^{R_c k/T}}{(2e^{R_c k/T}+r)^{2}} 
    \left(\frac{\delta R k}{T}\right)^{2}
    - \dfrac{re^{R_c k/T}(2e^{R_c k/T}-r)}{(2e^{R_c k/T}+r)^{3}} 
    \left(\frac{\delta R k}{T}\right)^{2} \left(\dfrac{mk}{T} \right)^{2} 
    \right\} \cr
    & \quad - \int_{k_\min}^{\infty} P_d(k) \, dk \,
    \left\{
    - \dfrac{re^{R_c k/T}(2e^{R_c k/T}-r)}{3(2e^{R_c k/T} + r)^{3}} 
    \left(\dfrac{\delta R k}{T} \right)^{3} \right\}
    + \mathcal{O}(m^{6}).
\end{align}
\begin{align}
    f
    & \simeq \dfrac{\langle k \rangle}{2 T} R_c^{2} 
    + C_{2}(\lambda) \left(\dfrac{m}{T} \right)^{2} 
    + C_{4}(\lambda) \left(\dfrac{m}{T} \right)^{4} 
    + C_{6}(\lambda) \left(\dfrac{m}{T} \right)^{6} + \mathcal{O}(m^{6}).
\end{align}

where the coefficients are given by
\begin{align}
    C_{2}(\lambda)
    &= \dfrac{T \langle k \rangle}{2} 
    - \int_{k_\min}^{\infty} k^{2} P_d(k) \, dk \, \dfrac{e^{R_c k/T}}{2e^{R_c k/T} + r}, \\[10pt]
    C_{4}(\lambda)
    &= \int_{k_\min}^{\infty} k^{4} P_d(k) \, dk \, 
    \dfrac{e^{R_c k/T}(4e^{R_c k/T}-r)}{12(2e^{R_c k/T} + r)^{2}} 
    - \frac{1}{2T} \frac{B^{2}(\lambda)}{A(\lambda)}, \\[10pt]
    C_{6}(\lambda)
    &= \int_{k_\min}^{\infty} k^{6} P_d(k) \, dk \, 
    \left[
    -\dfrac{e^{R_c k/T}(64e^{2R_c k/T}-26re^{R_c k/T}+r^{2})}{360(2e^{R_c k/T} + r)^{3}} \right] 
    - \frac{1}{T} \frac{B(\lambda)C_{1}(\lambda)}{A(\lambda)}
    - \frac{1}{2T} \frac{B^{2}(\lambda)C_{2}(\lambda)}{A^{2}(\lambda)} 
    - \frac{1}{3T} \frac{B^{3}(\lambda)C_{3}(\lambda)}{A^{3}(\lambda)}.
\end{align}

At the critical temperature where $C_{2}=0$, we identify the values of $(\lambda,T,r)$ satisfying $C_{4}=0$ and $C_{6}>0$. With four unknowns $(R_c,\lambda,T,r)$ and four equations, the system can be solved numerically, giving $\lambda_{u}\simeq 8.60$. This value coincides with the numerically determined boundary of the Goldilocks zone.

\section{Numerical Simulations}

\subsection{Network Generation}

Monte Carlo simulations were performed on annealed scale-free networks with $N = 10^6$ nodes, where the annealed approximation replaces the adjacency matrix with its degree-sequence average $\mathcal{A}_{ij} \rightarrow k_ik_j/(N\langle k\rangle)$, preserving degree heterogeneity while removing correlations and clustering. The degree sequence was generated from a continuous power-law distribution using inverse transform sampling:
\begin{equation}
    k_i = k_{\min} \left(1 - u_i\right)^{-1/(\lambda-1)},
\end{equation}
where $u_i$ are independent uniform random variables on $(0,1)$ and $k_{\min} = 1$ is the minimum degree. This deterministic sampling ensures exact realization of the target degree distribution. The degree exponents used were $\lambda = 3.9$ for the Ashkin--Teller model and $\lambda = 4.8$ for the Invisible Potts model.

\subsection{Simulation Protocol}

For both models, Monte Carlo simulations were performed on $\mathcal{H}_{\rm ann}$, with spin updates sampled according to the Boltzmann weight of $\mathcal{H}_{\rm ann}$. Each simulation consisted of:
\begin{itemize}
    \item \textbf{Equilibration:} $2.25 \times 10^4$ iterations
    \item \textbf{Sampling:} $1.225 \times 10^3$ iterations
\end{itemize}
All simulations started from the fully aligned (ferromagnetic) initial condition with all spins set to $+1$.

\subsection{Order Parameter Measurement}

The degree-resolved magnetization was computed by dividing the interval $1/k \in [0,1]$ into 20 equal-width bins. For each bin, we computed
\begin{equation}
    m_k = \frac{1}{N\langle k \rangle} \sum_{i \in \text{bin}} m_i k_i,
\end{equation}
such that $\sum_k m_k = m$. This binning in inverse degree ensures comparable statistical sampling across different degree classes.

\subsection{Error Estimation}

Error bars were determined from the standard deviation of the order parameter time series during the sampling period:
\begin{equation}
    \delta m = \sqrt{\frac{1}{n_s} \sum_{t=1}^{n_s} 
    \left(m(t) - \bar{m}\right)^2},
\end{equation}
where $n_s = 1.225 \times 10^3$ is the number of samples and $\bar{m}$ is the time-averaged order parameter.

\subsection{Parameter Ranges}

Simulations were performed over a range of temperatures $T$ and coupling parameters. For the Ashkin--Teller model, we used $x = 0.72$ and $1.23$ with corresponding temperature ranges near the transition points. For the Invisible Potts model, we used $r = 8.4$ with $\lambda = 4.8$.


\end{document}